\documentclass[sigconf]{acmart}

\usepackage{booktabs} %
\usepackage{amsmath}
\usepackage{algpseudocode}
\usepackage{subfigure}
\usepackage{tabularx}
\usepackage{bm}
\usepackage{xcolor}
\usepackage{cancel}
\usepackage{calc}
\usepackage{accents}
\usepackage{tikz}

\usepackage{tikz}
\usetikzlibrary{backgrounds}
\usetikzlibrary{calc}
\usetikzlibrary {shapes.geometric} 
\usetikzlibrary {fit}

\DeclareMathOperator*{\argmin}{arg\,min}

\DeclareMathOperator{\hess}{Hess}

\newcommand{\dbtilde}[1]{\accentset{\approx}{#1}}

\usepackage[ruled]{algorithm2e} %

\SetAlFnt{\small}
\SetAlCapFnt{\small}
\SetAlCapNameFnt{\small}
\SetAlCapHSkip{0pt}

\newcommand\todoCY[1]{{\color{black}{#1}}}

\copyrightyear{2023} 
\acmYear{2023} 
\setcopyright{acmlicensed}\acmConference[SA Conference Papers '23]{SIGGRAPH Asia 2023 Conference Papers}{December 12--15, 2023}{Sydney, NSW, Australia}
\acmBooktitle{SIGGRAPH Asia 2023 Conference Papers (SA Conference Papers '23), December 12--15, 2023, Sydney, NSW, Australia}
\acmPrice{15.00}
\acmDOI{10.1145/3610548.3618158}
\acmISBN{979-8-4007-0315-7/23/12}

\begin{document}
\title{LiCROM: Linear-Subspace Continuous Reduced Order Modeling with Neural Fields}

\author{Yue Chang}
\orcid{0000-0002-2587-827X}
\affiliation{%
  \institution{University of Toronto}
  \country{Canada}}
\email{changyue.chang@mail.utoronto.ca}

\author{Peter Yichen Chen}
\orcid{0000-0003-1919-5437}
\affiliation{%
  \institution{MIT CSAIL}
  \country{USA}}
\authornote{Corresponding authors (e-mail: pyc@csail.mit.edu, eitan@cs.toronto.edu).}
\email{pyc@csail.mit.edu}

\author{Zhecheng Wang}
\orcid{0000-0003-4989-6971}
\affiliation{%
  \institution{ University of Toronto}
  \country{Canada}}
\email{zhecheng@cs.toronto.edu}

\author{Maurizio M. Chiaramonte}
\orcid{0000-0002-2529-3159}
\affiliation{%
  \institution{Meta Reality Labs Research}
  \country{USA}}
\email{mchiaram@meta.com}

\author{Kevin Carlberg}
\orcid{0000-0001-8313-7720}
\affiliation{%
  \institution{Meta Reality Labs Research}
  \country{USA}}
\email{carlberg@meta.com}

\author{Eitan Grinspun}
\orcid{0000-0003-4460-7747}
\affiliation{%
  \institution{ University of Toronto}
  \country{Canada}}
\email{eitan@cs.toronto.edu}

\renewcommand{\shortauthors}{Y. Chang, P. Chen, Z. Wang, M. Chiaramonte, K. Carlberg, E. Grinspun}
\authornotemark[1]

\begin{abstract}
Linear reduced-order modeling (ROM) simplifies complex simulations by approximating the behavior of a system using a simplified kinematic representation. Typically, ROM
is trained on input simulations created with a specific spatial discretization, 
and then serves to accelerate simulations with the same discretization. 
This discretization-dependence is restrictive. 

Becoming independent of a specific discretization would provide flexibility to mix and match  mesh resolutions, connectivity, and type (tetrahedral, hexahedral) in training data; to 
accelerate simulations with novel discretizations unseen during training; 
and to accelerate adaptive simulations that temporally or parametrically change
the discretization. 

We present a flexible, discretization-independent approach to reduced-order modeling. 
Like traditional ROM, we represent the configuration as a linear combination of displacement
fields. Unlike traditional ROM, our displacement fields are continuous maps from every point on the reference domain to a corresponding displacement vector; these maps are
represented as implicit neural fields.

With linear continuous ROM (LiCROM), our training set can include multiple geometries undergoing multiple loading conditions, independent of their discretization. This opens the door to novel applications of reduced order modeling. We can now accelerate simulations that modify the geometry at runtime, for instance via cutting, hole punching, and even swapping the entire mesh. We can also accelerate simulations of geometries unseen during training. 
We demonstrate one-shot generalization, training on a single geometry and subsequently simulating various unseen
geometries.
\end{abstract}




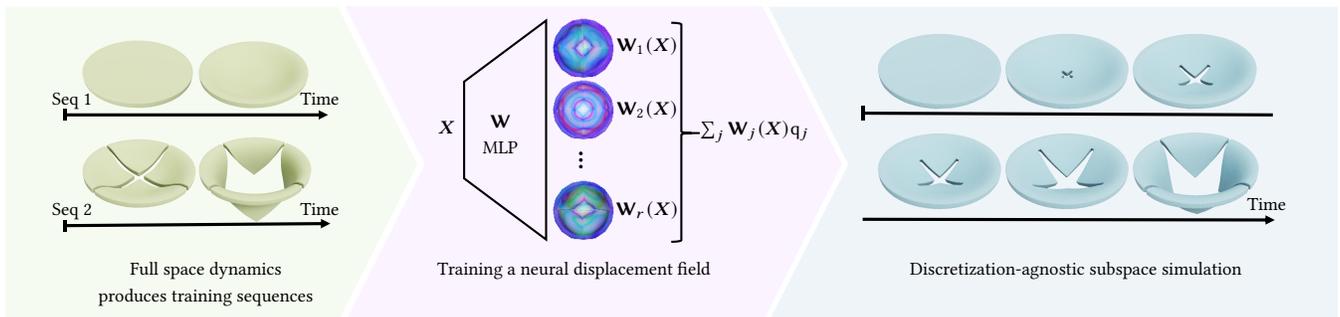
\begin{teaserfigure}
\centering
\begin{tikzpicture}[x=0.5\textwidth, y=0.5\textwidth]
\node[anchor=south] (image) at (0,0) {
\hspace{-0.08cm}\includegraphics[width=17.7cm]{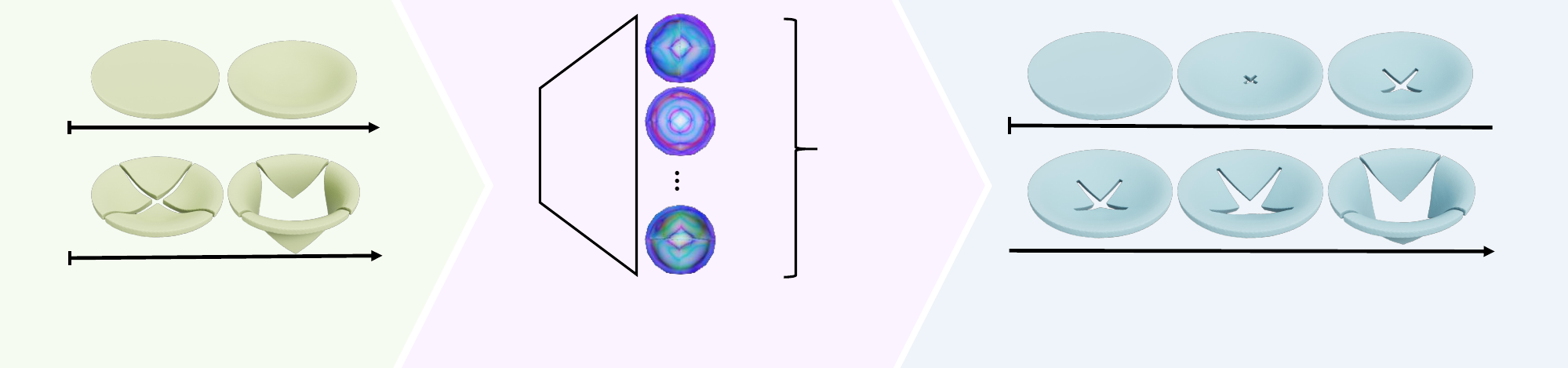}
};
\node at ([shift={(-0.7, 0.11)}]image.south)[below] 
  {\footnotesize Full space dynamics };  
  \node at ([shift={(-0.7, 0.07)}]image.south)[below] 
  {\footnotesize produces training sequences};  
\node at ([shift={(0.6, 0.11)}]image.south)[below] 
  {\footnotesize Discretization-agnostic subspace simulation};  
\node at ([shift={(-0.34, 0.32)}]image.south)[below] 
  {\footnotesize $\bm{X}$};  
\node at ([shift={(-0.26, 0.29)}]image.south)[below] 
  {\footnotesize MLP};  
\node at ([shift={(-0.26, 0.33)}]image.south)[below] 
  {\footnotesize $\bm{\mathsf{W}}$};  
\node at ([shift={(-0.04, 0.445)}]image.south)[below] 
  {\footnotesize $\bm{\mathsf{W}}_1(\bm{X})$};
\node at ([shift={(-0.04, 0.35)}]image.south)[below] 
  {\footnotesize $\bm{\mathsf{W}}_2(\bm{X})$};
\node at ([shift={(-0.04, 0.2)}]image.south)[below] 
  {\footnotesize $\bm{\mathsf{W}}_r(\bm{X})$};
\node at ([shift={(0.12, 0.32)}]image.south)[below] 
  {\footnotesize $\sum_j{\bm{\mathsf{W}}_j(\bm{X})}\mathsf{q}_j$};
\node at ([shift={(-0.15, 0.11)}]image.south)[below] 
  {\footnotesize Training a neural displacement field};  
  \node at ([shift={(0.885, 0.205)}]image.south)[below] 
  {\footnotesize Time};
\node at ([shift={(-0.53, 0.363)}]image.south)[below] 
  {\footnotesize Time};
\node at ([shift={(-0.53, 0.198)}]image.south)[below] 
  {\footnotesize Time};
\node at ([shift={(-0.9, 0.363)}]image.south)[below] 
  {\footnotesize Seq 1};
\node at ([shift={(-0.9, 0.198)}]image.south)[below] 
  {\footnotesize Seq 2};
\end{tikzpicture}
\caption{Starting with training simulations (\emph{left}), we
train a linear subspace of neural displacement fields (\emph{center}).  Each such field $\bm{\mathsf{W}}_j$ 
maps material position $\bm{X}$ to displacement $\bm{\mathsf{W}}_j(\bm{X})$. Because the map is continuous, it effectively forgets the discretizations used in the training simulations.  Consequently,
the resulting linear subspace $\sum_j{\bm{\mathsf{W}}_j(\bm{X})}\mathsf{q}_j$ is
ideally suited for fast simulations of scenarios that benefit from adaptive discretization, such as 
progressive cutting and topology changes (\emph{right}).}
\end{teaserfigure}

\maketitle

\section{Introduction}

Reduced-order modeling (ROM) using linear subspaces to approximate the solution space can accelerate deformable object simulations by orders of magnitude. 
The idea is to generate a number of simulated trajectory exemplars, and then identify a low-dimensional basis that approximates the exemplar displacements. We then compute dynamics by evolving only the small number of coefficients of this basis, known as \emph{reduced coordinates} or \emph{latent variables.}

Classical approaches to ROM assume that the input exemplars and output dynamics are all represented by a given spatial discretization, say a mesh of the domain $\Omega \subset \mathbb{R}^3$. This reliance on a specific  discretization can be restrictive.

Being untethered from a specific discretization is desirable when input exemplars are produced using different meshes (e.g., different connectivity or resolution); simulation outputs are desired for various meshes; we wish to produce simulation output that temporally or parametrically adapts the mesh to suit the deformation (e.g., dynamic remeshing, arbitrary Lagrangian--Eulerian simulation).

Indeed, variations need not be limited to mesh connectivity and resolution: perhaps we want to vary the mesh \emph{type} (e.g., quad versus tetrahedral meshes) or even the discretization type (e.g., mesh, point sets with generalized moving least squares, radial basis functions, spectral discretizations).

We present such a discretization-agnostic approach to reduced order modeling. 
Our approach retains the linearity of the subspace of common ROM approaches,
but substitutes 
the discrete representation of each displacement basis field with its continuous analogue. 

To make things concrete, consider a simple classical ROM approach tied to a 
mesh with $n$ vertices. We denote the time-varying displacement of the mesh from its reference configuration 
by $\bm{\overline{u}}(t)$ with $\bm{\overline{u}}: \mathcal T \rightarrow \mathbb{R}^{3n}$, where $\mathcal T (\subseteq \mathbb{R})$
denotes the temporal domain.
We will place a bar (e.g., $\bm{\overline{u}}(t)$) over those quantities
that depend on spatial discretization, i.e., those with an index ranging over $1\ldots n$.

In classical ROM, we approximate the time-varying displacement of the mesh
as a linear combination $\bm{\overline{u}}(t) \approx \overline{\mathsf{U}} \mathsf{q}(t)$
of some $r \ll n$ dimensional, time-independent basis
$\overline{\mathsf{U}}$, 
where
$\mathsf{q}(t) : \mathcal T  \rightarrow \mathcal{Q}$ 
is the reduced or \emph{latent} trajectory in the latent subspace $\mathcal{Q} \subset \mathbb{R}^r$,
and $\overline{\mathsf{U}} \in \mathcal{M}_{3n \times r}(\mathbb{R})$
is typically found via Proper Orthogonal Decomposition\footnote{POD is also known 
as the Karhunen--Lo\`eve transform 
and is closely related to Principal Component Analysis (PCA).}
(POD) over a training set of simulation data (temporal sequences of displacement fields);
$\mathcal M_{m\times n}(A)$ denotes the set of $m\times n$ matrices over the field $A$.
Each column $\overline{\mathsf{U}}_k$ is a particular \emph{discrete} displacement field over the 
$n$ vertices; the mutually orthogonal 
columns $\{\overline{\mathsf{U}}_1\ldots \overline{\mathsf{U}}_r\}$ form the basis for the \emph{discrete} displacement subspace. We will use the sans serif typeface ($\overline{\mathsf{U}}$, $\mathsf{q}$) to denote quantities that depend on the subspace dimension $r$.

Now here is the crux of the matter: the discrete ``architecture'' of $\overline{\mathsf{U}}$ is immutably anchored to the initial discretization. 
The $j$th row of $\overline{\mathsf{U}}$ is the basis for the $j$th degree of freedom,
where $1\leq j \leq n$. 
Indeed, for a mesh discretization, the temporal evolution of the three degrees of freedom associated with $i$th vertex is given by
\begin{align}\label{eq:W-discrete}
\bm{\overline{u}}_i(t) = {\overline{\bm{\mathsf{W}}}_i} \mathsf{q}(t) \ ,
\end{align}
where  $\overline{\bm{\mathsf{W}}}_i \in \mathcal{M}_{1 \times r}(\mathbb{R}^3)$  is a $1 \times r$ matrix (a row vector) of $\mathbb{R}^3$-valued coefficients, i.e., one displacement vector per 
each of the $r$ subspace modes. The $3 \times r$ coefficients of $\overline{\bm{\mathsf{W}}}_i$ are drawn from those $3$ rows of $\overline{\mathsf{U}}$ corresponding to vertex $i$. 
(We will use boldface to denote $\mathbb{R}^3$-valued entries.)

Stacking the row vectors $\overline{\bm{\mathsf{W}}}_i$ of all vertices gives $\overline{\bm{\mathsf{W}}} \in \mathcal{M}_{n \times r}(\mathbb{R}^3)$, an $n \times r$ matrix with $\mathbb{R}^3$-valued entries, mapping
$\bm{\overline{u}}(t) = \overline{\bm{\mathsf{W}}} \mathsf{q}(t)$.
Essentially, $\overline{\bm{\mathsf{W}}}$ encodes the time-invariant linear mapping from the latent configuration $\mathsf{q}(t)$ to the full space displacements $\bm{\overline{u}}(t)$.

We are nearly ready for our novel step, the transition to the smooth setting. We view $\overline{\bm{\mathsf{W}}} = i \mapsto \overline{\bm{\mathsf{W}}}_i :  \{1, \ldots, n\} \rightarrow \mathcal{M}_{1 \times r}(\mathbb{R}^3)$
as a map from the vertex index to the row vector of subspace weights.
This is a discrete map, and that is what we will now make smooth.

\begin{figure}
    \centering
    \begin{tikzpicture}[x=0.5\textwidth, y=0.5\textwidth]
        \edef\radius{0.1}
        \edef\perturbation{0.01}
        \edef\shift{0.4}
        \draw [red, thick, fill=black!10!white,domain=0:360, samples=360] 
        plot ( { \shift + (\radius+ cos(\x*10)*\perturbation)*cos(\x) } , { (\radius + cos(\x*10)*\perturbation)*sin(\x)} );
        \draw [red, thick, fill=black!10!white,domain=0:360, samples=360] 
        plot ( { (\radius+ cos(\x*4)*\perturbation)*cos(\x) } , { (\radius + cos(\x*4)*\perturbation)*sin(\x)} );
        \coordinate (XREF) at ( 0, {-\radius/2} );
        \coordinate (XDEF) at ({\shift },{-\radius/2} );
        \node at (XREF) { $\bm{X}$ \textbullet};
        \node at (XDEF) {\textbullet $\bm{x}$};
        \draw [latex-, thick, shorten >=0.1cm, shorten <=0.1cm] ({-\radius+\shift}, {\radius/2})  to[out=135,in=45]node[pos=0.5,above]{$\bm{X} + \bm{u}(\bm{X}, t)$} (\radius, {\radius/2}) ;
        \draw [-latex, thick, shorten >=0.25cm, shorten <=0.25cm] (XREF)  to node[pos=0.5,above]{$\bm{u}(\bm{X},t)$} (XDEF) ;
        \node at ({\shift},{\radius/2})  {$\Omega_t$};
        \node at ({0},{\radius/2})  {$\Omega$};
    \end{tikzpicture}
    \vspace{-5pt}
    \caption{\emph{Deformation of an elastic body.} The reference domain $\Omega$ and the deformed domain $\Omega_t$ at time $t$ 
    are related by the deformation mapping $\bm{X} \mapsto \bm{x}(\bm{X},t) = \bm{X} + \bm{u}(\bm{X},t)$: 
    each deformed point $\bm{x}(\bm{X},t)$ is displaced by $\bm{u}(\bm{X},t)$
    relative to the reference point $\bm{X}$.}
    \vspace{-15pt}
    \label{fig:domains-and-maps}
\end{figure}
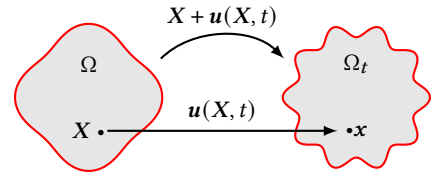

In lieu of the discrete map $\overline{\bm{\mathsf{W}}}$,
we propose to instead use a continuous map $\bm{\mathsf{W}} = \bm{X} \mapsto \bm{\mathsf{W}}(\bm{X}): \Omega \rightarrow \mathcal{M}_{1 \times r}(\mathbb{R}^3)$ taking
a point $\bm{X} \in \Omega$ in the reference domain to its subspace weights, 
so that
\begin{align} \label{eq:W-continuous}
\bm{u}(\bm{X},t) = \bm{\mathsf{W}}(\bm{X}) \mathsf{q}(t) \ .
\end{align}
Comparing to \eqref{eq:W-discrete}, the discretization-dependent \emph{discrete} index $i$ is replaced by a discretization-independent \emph{continuous}
reference point $\bm{X}$ (see Fig.~\ref{fig:domains-and-maps}). 
The time-varying spatially-varying displacement field $\bm{u}(\bm{X},t)$
is a linear combination of spatially-varying, time-invariant 
basis of displacement fields, whose time-varying, spatially-invariant 
weights are given by $\mathsf{q}(t)$.

To aid in intuition, we can also compare the columns of $\overline{\mathsf{U}}$, 
$\overline{\bm{\mathsf{W}}}$, and $\bm{\mathsf{W}}$. In all cases, the $k$th column is
a representation of a particular displacement---a basis element of the approximating subspace---as a field over
the entire domain $\Omega$; the distinction is that $\bm{\mathsf{W}}_k$ is a continuous
field, whereas the others are discrete column vectors.

\begin{figure}
\centering

\begin{tikzpicture}[x=0.5\textwidth, y=0.5\textwidth]
\node[anchor=south] (image) at (0,0) {
\includegraphics[width=8cm]{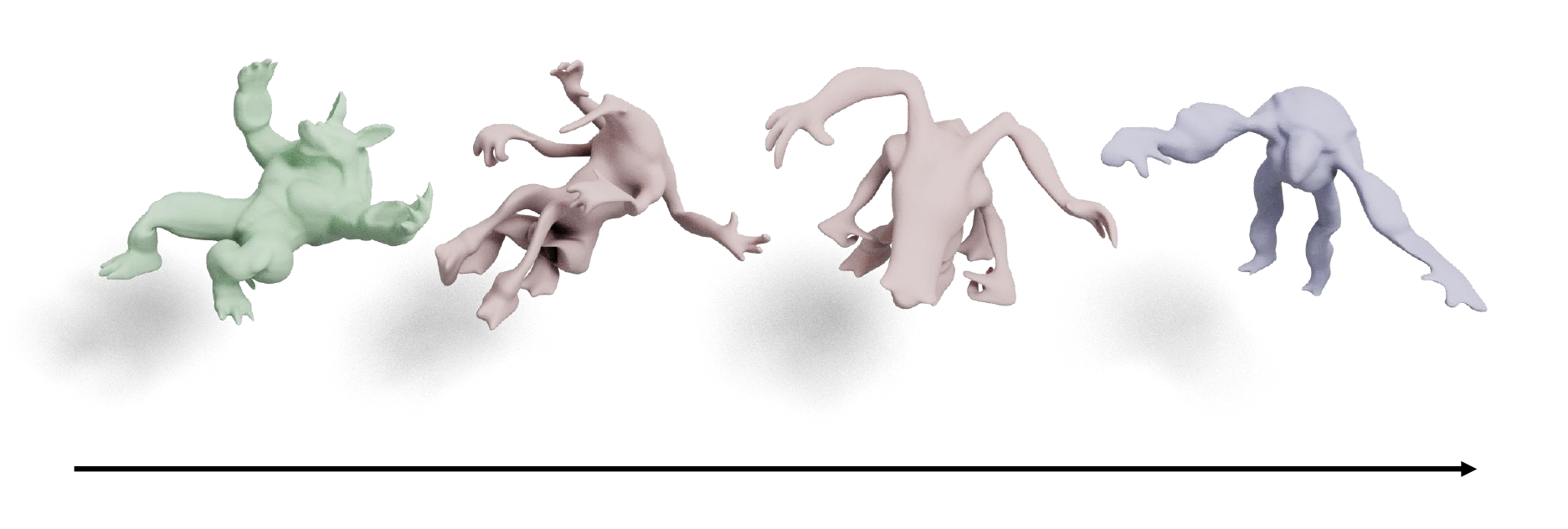}
};
\node at ([shift={(0.17, 0.32)}]image.south)[below] 
  {\footnotesize Change character};
\node at ([shift={(-0.2, 0.32)}]image.south)[below] 
  {\footnotesize Change character};  
\node at ([shift={(-0.0, 0.1)}]image.south)[below] 
  {\footnotesize Time stepping in reduced space};  
\end{tikzpicture}

\vspace{-15pt}
\caption{
\emph{Interactive manipulation and one-shot generalization.}
Training a neural basis on deformations of the Armadillo,
our application allows the user to interactively tug at the 
geometry. Unlike discretization-dependent reduction techniques,
we can easily substitute the geometry. We compute the 
latent dynamics on three meshes not seen during training.
While the kinematics are defined by the training set, the physical
response is defined by the geometry of the current mesh, as 
evident in details, such as the wobbling of the stick arms.
Frame rate: 30 frames per second.
Full space time step cost: $335$ms; reduced: $6$ms.
Hardware: Intel Core i7-10750H.
\label{fig:interactive}
}
\centering
\end{figure}
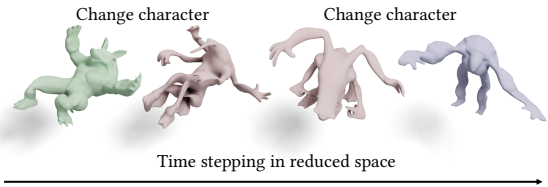

Equation \ref{eq:W-continuous} is the basis for \emph{linear continuous ROM} (LiCROM).
Now the training set can span multiple discretizations of the same geometry, or even multiple geometries. This facilitates and broadens the applicability of reduced order modeling: As we
will show, with LiCROM we can compute latent dynamics on geometries unseen during training;  simulations that modify the geometry at runtime via cutting, hole punching, or 
swapping the entire mesh (Fig. \ref{fig:interactive}), without re-initializing the reduced coordinates. 

\section{Related Work}
\paragraph{Linear reduced-order modeling} Model-reduction techniques
\citep{benner2015survey} have proven to be a powerful tool for 
enabling high-fidelity models 
to be run in real-time. 
\todoCY{They have been successfully applied to problems in many fields, such as fluid dynamics \citep{hall2000proper,
willcox2002balanced,bergmann2005optimal,lieu2006reduced,carlberg2013gnat,mainini2015surrogate,carlbergGalDiscOpt,treuille2006model,kim2019deep,kim2013subspace,wiewel2019latent}, solid mechanics
\citep{Barbic:RTSI:2005,james2006precomputed,barbivc2011real,yang2015expediting,an2008optimizing,kim2009skipping,xu2015interactive}, secondary motion for rigged animation~\citep{Xu:2016:PSD, benchekroun2023fast} and robotics
\citep{katzschmann2019dynamically,tan2020realtime}. 
}

\todoCY{
Typically, the reduced space is learned from training exemplars \citep{berkooz1993proper,Barbic:RTSI:2005, Fulton:LSD:2018}, or identified in a ``data-free'' manner from energetic first principles \citep{pentland1989good,shabana2012vibration, yang2015expediting, sharp2023data}. 
``Online'' approaches update the basis at runtime based on the observed trajectory~\citep{Ryckelynck:2005:Priori, kim2009skipping, Mukherjee:2016:IDSR}; a related approach is to interpolate between precomputed bases~\citep{Xu:2016:PSD}. 
We learn a fixed basis from simulated exemplars.}

Most model-reduction methods employ a linear-subspace approximation for the kinematics. Such approximations are accurate for problems displaying a rapidly-decaying Kolmogorov $n$-width \cite{pinkus2012n}. 
However, nearly all of these operate with a discrete representation; those that do operate with the continuous representation (e.g., reduced-basis methods) are intrinsically tied to an underlying spatial discretization scheme.
There have been a few methods that applied nonlinear kinematic approximations, which we will discuss below. Crucially, most of these also operate on a \emph{discrete} representation, with the exception of CROM~\cite{CHEN:CROM:2023,CHEN:CROM-MPM:2023}, which has been applied to the material point method
and to various partial differential equations.

We fill the gap in the literature by developing the first \textit{linear} kinematic approximation that is also independent of any spatial discretization.

\paragraph{Deep-learning-based reduced-order modeling}
\citeN{Lee2018rom} introduced the first framework utilizing autoencoders to capture nonlinear manifolds. \citeN{Fulton:LSD:2018} extended this idea, combining it with POD for deformable solid dynamics. In a complementary approach, \citeN{Shen:HOD:2021} used nonlinear autoencoders to efficiently execute Hessian-based latent space dynamics by accurately computing high-order neural network derivatives. Furthermore, \citeN{Romero:LCCHSD:2021} introduced contact-induced deformation correction with linear subspace modes. Meanwhile, \citeN{Luo:2020:NNWarp} focused on displacement correction, aiming to transform linear elastic responses into more complex constitutive ones.

\vspace*{-0.08in}

\paragraph{Discretization-independent representations} 

Recently, implicit neural representations have become an exciting area of exploration in many fields, including shape modeling~\cite{CHEN:IM-NET:2019, Park:DeepSDF:2019}, 3D reconstruction~\cite{Mescheder:ON:2019, Mildenhall:NeRF:2021}, image representation and generation~\cite{Shaham:SAPN:2021,Skorokhodov:AGCI:2021, Chen:LCIR:2021}, and PDE-constrained problems~\cite{Raissi:PINN:2019,zehnder2021ntopo,yang2021geometry,chen2022implicit}.  

\citeN{Aigerman:NJF:2022} proposed a framework to accurately predict piecewise linear mappings of arbitrary meshes using a neural network. It works with heterogeneous collections of meshes without requiring a shared triangulation. 
Others aim to learn the latent space representation of continuous vectors: \citeN{CHEN:CROM-MPM:2023} proposed a model reduction method for material point method, while \citeN{CHEN:CROM:2023} and \citeN{Pan:NIF:2023} learned a discretization-agnostic latent space for PDEs. 
To the best of our knowledge, the prototypical
factored structure of linear ROM,
$\bm{\mathsf{W}}(\bm{X}) \mathsf{q}(t)$, 
has not been considered in the context of 
continuous discretization-independent representations for model reduction.

\section{Discretization-blind subspace learning}
We train LiCROM over an observed trajectory of a deformable object.
To simplify notation,
assume one trajectory sampled at instances $\{t^1,\ldots,t^m\}$, 
although the approach trivially generalizes
to sampling multiple trajectories or multiple objects with parallel trajectories.

Let $\mathbb{X} = \{
(\tilde{\bm{X}}^1,
\tilde{\bm{u}}^1),\ldots,
(\tilde{\bm{X}}^m,
\tilde{\bm{u}}^m)
\}$ be
the training set, where $(\tilde{\bm{X}}^j,\tilde{\bm{u}}^j)$ collects 
observations of the displacement field at time $t^j$.
In particular, 
$\tilde{\bm{u}}^j = 
\{\,\bm{u}^j_1,\,\bm{u}^j_2,\,\ldots\}\subset\mathbb{R}^3$
consists of a finite number of observations 
$\bm{u}^j_i \equiv \bm{u}(\bm{X}^j_i,t^j)$
of the displacement field at reference positions 
$\tilde{\bm{X}}^j \equiv 
\{\,\bm{X}^j_1,\,\bm{X}^j_2,\,\ldots\}$.
We do not assume a consistent structure between point clouds, i.e., 
the sample positions $\bm{X}^j_i$ and $\bm{X}^{j+1}_i$ need not be equal, nor 
the sample counts $|\tilde{\bm{X}}^j|$ and $|\tilde{\bm{X}}^{j+1}|$.

We seek a low-dimensional subspace 
that spans all the observed fields $(\tilde{\bm{X}}^j,\tilde{\bm{u}}^j)$.
In particular, we seek a projection 
$\mathsf{P}:\ (\tilde{\bm{X}}^j,\tilde{\bm{u}}^j) \mapsto \mathsf{q}^j
\in \mathcal{Q}$,
and a corresponding basis $\bm{\mathsf{W}}$ (independent of $j$) such that
\begin{align} \label{eq:training-overview}
\bm{\mathsf{W}}(\bm{X}_i) \mathsf{P} (\tilde{\bm{X}}^j,\tilde{\bm{u}}^j) 
\approx \bm{u}^j_i \ ,\quad 
\forall\, (\tilde{\bm{X}}^j,\tilde{\bm{u}}^j) \, \in\, \mathbb{X}\ ,\quad 
\forall\, \bm{X}_i \, \in\, \tilde{\bm{X}}^j\ . 
\end{align}

\begin{figure}

\begin{tikzpicture}
\node[name=P, trapezium, rotate=270, trapezium angle=60, minimum width=4cm,  draw] {};
\draw (P.center) node[align=center] {$\mathsf{P}$ \\ \small PointNet};
\draw[<-,>=latex] (P.south) -- ++(-.5,0) node[left,name=PointNetInput] {$\underbrace{\begin{bmatrix}
(\tilde{\bm{X}}_1, \tilde{\bm{u}}_1(t)) \\
(\tilde{\bm{X}}_2, \tilde{\bm{u}}_2(t)) \\
\vdots \\
(\tilde{\bm{X}}_N, \tilde{\bm{u}}_N(t)) \\
\end{bmatrix}}_{\mathbb{X}}$} ;
\draw[->,>=latex]  (P.north) -- node[above,name=ReducedConfig] {$\mathsf{q}(t)$} ++(+1.5,0) node[name=product,right] {\huge $\otimes$}; 

\draw (product) ++(0,+2) node[name=W, trapezium, trapezium angle=70, minimum width=1cm, minimum height=1cm, draw] {};
\draw (W.center) node[align=center] {$\bm{\mathsf{W}}$ \\ \small MLP};
\draw[<-,>=latex] (W.north) -- ++(0,+.5) node[above,name=ReferencePoint] {$\bm{X}_i$};
\draw[->,>=latex]  (W.south) -- node[right,name=WXi] {$\bm{\mathsf{W}}(\bm{X}_i)$} (product)[above];

\draw (product)[right] ++(0.2,0) node[align=center,right] {$\stackrel{\textrm{matrix}}{\footnotesize \textrm{product}}$};

\draw[->,>=latex] (product)[below] -- node[right,name=WXq] {$\bm{\mathsf{W}}(\bm{X})\mathsf{q}(t)$} ++(0,-2) node[below,name=output] {$\bm{u}(\bm{X},t)$};

\begin{scope}[on background layer]

\node[name=TrainingRegion,fit=(PointNetInput) (P.bottom left corner) (P.bottom right corner) (P.top side),fill=blue!10] {};

\node[below] at (TrainingRegion.south) {Used only during training};

\node[name=CachingRegion,fit=(W.bottom left corner) (W.bottom right corner) (ReferencePoint),fill=green!10] {}; 

\node[left,align=right] at ([shift={(0,0.5)}]CachingRegion.west) {Cached for \\ all cubature points};

\node[name=MainLoop,fit=(product) (WXi) (WXq) (output) (ReducedConfig),fill=red!10] {}; 

\node[below,align=center] at ([shift={(0,0)}]MainLoop.south) {Subspace dynamics loop};

\end{scope}

\end{tikzpicture}
\vspace{-10pt}
\caption{\emph{Network architecture.} To learn the subspace, 
the PointNet encoder $\mathsf{P}(\bm{X},\bm{u})$ 
and the 5-layer multilayer perceptron $\bm{\mathsf{W}}(\bm{X})$ are optimized over the
training set $\mathbb{X}$. This yields a compact, factored kinematic space where the 
displacement field $\bm{u}(\bm{X},t)$ is a $\mathsf{q}(t)$-weighted linear combination of 
$r$ precomputed time-invariant continuous displacement fields $\{\bm{\mathsf{W}}_1(\bm{X})\ldots\bm{\mathsf{W}}_r(\bm{X})\}$. 
During subspace simulation, the PointNet encoder is no longer required: 
it has served its
purpose, finding an embedding for $\mathsf{q}(t)$. The 
evaluation of $\bm{\mathsf{W}}(\bm{X})$ at cubature points $\{\bm{X}_i\}$
is performed once when the cubature scheme is established, and cached for reuse.
Subsequently, the subspace simulation inner loop requires only the usual 
matrix-vector product  
$\bm{\mathsf{W}}(\bm{X}_i)\mathsf{q}$ that is the cornerstone of linear subspace methods.
}\label{fig:network}
\vspace{-15pt}
\centering
\end{figure}
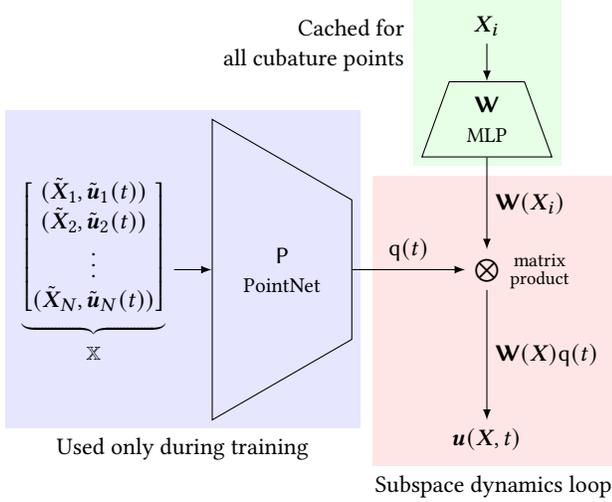

We adopt a parametric form for $\mathsf{P}$ and $\bm{\mathsf{W}}$,
in particular a PointNet encoder~\cite{Charles:PointNet:2017}
and neural implicit field~\cite{Mescheder:ON:2019,Park:DeepSDF:2019}, respectively,
and optimize the parameters to minimize the squared norm residual 
of \eqref{eq:training-overview},
as depicted in Fig. \ref{fig:network}.

\subsection{Training Method}

In our experiments, we produced the training set using simulations based on 
tetrahedral mesh discretizations. However, observe that the network does not directly 
``know'' that the input was generated by a mesh, only that a sampled displacement field was generated
somehow. The network aims to find a reduced basis that can reconstruct all observed
displacement fields, without consideration for loading, boundary conditions, geometry, or discretization.

To produce our training set, we first generate a volumetric tetrahedal mesh for each geometry using TetWild \cite{Hu:TMW:2018}, and then execute the desired full-space simulation using a CPU-based  \emph{taichi}~\cite{HU:TAICHI:2019}
implementation that closely follows the default implicit FEM integrator in \emph{warp}~\cite{warp2022}.
We repeat this process
to produce a set of simulation results. Our implementation uses the same cardinality $\tilde{n} = |(\tilde{\bm{X}}^j,\tilde{\bm{u}}^j)|$ for the randomly sampled vertex-based displacements of each animation frame, which simplifies batch processing in PyTorch. We found that training the PointNet encoder $\mathsf{P}$ can be expensive when ($\tilde{n} > 5000$), yet using a large cardinality is helpful for coverage of the domain. Therefore, we further subsample $\dbtilde{n} < \tilde{n}$ vertices for the PointNet encoder. We determine the parameters for $\mathsf{P}$ and $\bm{\mathsf{W}}$ by minimizing the $L_2$ reconstruction loss 
\begin{align}
\mathcal{L}=\sum_{j=1}^{m} \sum_{i=1}^{\tilde{n}}\ 
\left\| \bm{\mathsf{W}}(\bm{X}_i) \mathsf{P} \circ \mathsf{S}_{\dbtilde{n}} (\tilde{\bm{X}}^j,\tilde{\bm{u}}^j) 
- \bm{u}^j_i\right \|_2 \ , 
\end{align}
where $\mathsf{S}_{\dbtilde{n}}$ is the subsampling operator. We used $\dbtilde{n}=2500$ for all examples.

\paragraph*{PointNet architecture}
The PointNet encoder $\mathsf{P}$ is invariant under permutation of input points, a desirable feature 
for our unordered sets. A standard PointNet is also
invariant under input transformations due to its input stage feature-transform net; we removed
this stage since latent space variables are not invariant under transformations of \emph{displacements}. The input to the PointNet is an unordered set of points $(\bm{X}_i,\bm{u}_i) \in \mathbb{R}^3 \times \mathbb{R}^3 \equiv \mathbb{R}^6$ and the output is $\mathsf{q}$.

\paragraph*{Neural field architecture}
The architecture for the neural field $\bm{\mathsf{W}}$ is a 5-layer multilayer perceptron (MLP) of width 60  
with ELU~\cite{clevert:elu:2016} activation functions. We used this architecture for all
presented examples, however, we found that alternatives such as SIREN~\cite{Sitzmann:SIREN:2020} can also generate good results.

\paragraph*{Learning network parameters}
We use PyTorch Lightning to implement the entire training pipeline \cite{Falcon:PyTorchLightning:2019}. We adopted the Adam optimizer~\cite{kingma:adam:2017} and apply Xavier initialization. We train the model for $3750$ epochs with a base learning rate of $\textrm{lr} = 10^{-3}$. After the first $1250$ epochs, we divide the learning rate by $5$, then we further divide it by $10$ after another $1250$ epochs. We used a batch size 16 for the network's input, so the batch size is $16 \cdot \tilde{n}$ for $W$.

\section{Dynamics via Implicit integration}

We formulate an implicit timestep in the framework of optimization time integrators
\cite{Stuart:1996:DynamicalSA,Pan:2015:SDSR,Martin:2011:EBEM}, which were
recently used for latent space dynamics by \citeN{Fulton:LSD:2018}. 
The configuration $\mathsf{q}$ at the end of the $(j+1)$th time step minimizes
\begin{align}\label{eq:reduced_space_energy_ours_smooth}
     E(\mathsf{q}) = \int_{\bm{X} \in \Omega} \frac{1}{2h^2}\big\|\bm{\mathsf{W}}(\bm{X})\mathsf{q}-\bm{u}_{\textrm{pred}}\big\|_g + \Psi(\bm{X}+\bm{\mathsf{W}}(\bm{X})\mathsf{q}) \ \textrm{dVol} \ ,
\end{align}
where $h$ is the duration of the time step,
$g$ is the kinetic energy\footnote{
The kinetic energy norm $\|\bm{v}(\bm{X})\|_g 
= \int_{\Omega} \frac{1}{2}\rho(\bm{X})\bm{v}(\bm{X})^2 \mathrm{dVol}$,
where $\rho(\bm{X})$ is the mass density.
} norm, 
and $\Psi(\bm{x})$ is the elastic energy density, in our 
implementation stable neohookean~\cite{Smith:stable-neohookean:2018}. The explicit predictor
for the $(j+1)$th time step
\begin{align}
\bm{u}_{\textrm{pred}}^{j+1} = \bm{u}^j + h \bm{v}^n + h^2 M^{-1} \bm{f}_{\textrm{{ext}}} 
\end{align}
requires the full-space velocity given by the finite difference
\begin{align}
\bm{v}^n=\frac{\bm{u}^n-\bm{u}^{n-1}}{h}=\bm{\mathsf{W}}(\bm{X})\frac{\mathsf{q}^n-\mathsf{q}^{n-1}}{h} = \bm{\mathsf{W}}(\bm{X}) \dot{\mathsf{q}}^n \ ,
\end{align}
where (by linearity of the subspace) $\dot{\mathsf{q}}^n=(\mathsf{q}^n-\mathsf{q}^{n-1})/h$.

We approximate the domain integral \eqref{eq:reduced_space_energy_ours_smooth} via 
cubature
\begin{align}\label{eq:reduced_space_energy_ours}
     E(\mathsf{q}) \approx \sum_i \frac{w_i}{2h^2}\|\bm{\mathsf{W}}(\bm{X}_i)\mathsf{q}-\bm{u}_{\textrm{pred}}\|_g + w_i \Psi(\bm{X}_i+\bm{\mathsf{W}}(\bm{X}_i)\mathsf{q}) \ ,
\end{align}
where $w_i$ is the weighting of the $i$th cubature point $\bm{X}_i$.
Our implementation performs the cubature and energy density computation using a mesh,
motivated by the readily available methods for volumetric 
deformables~\cite{An:Cubature:2008}, 
although the mathematics are not tied to mesh-based cubature.

Regardless, the cubature mesh is not and need not be tied to the representation
of the training data. Furthermore, the
cubature mesh need not be the same across time steps, since the state is 
carried across time steps by the latent configuration $\mathsf{q}$. 
The cubature should be chosen to adequately control the approximation
\eqref{eq:reduced_space_energy_ours} and to enforce
the essential boundary conditions.

This freedom makes
scenarios that have connectivity changes (e.g., fracture, cutting), and
topology changes (e.g., punching out a hole, growth of voids) refreshingly
trivial: we simply choose an appropriate cubature scheme for 
the next time step. For instance, if a hole is instantaneously punched out,
we simply refrain from integrating over the excised domain, 
by switching to a cubature mesh that reflects the revised topology
and revised boundary conditions. 

An alternative to switching the mesh would be to skip
cubature points that lie in the void. The key point is that there is
a lot of freedom in the approach---even across time steps---to integrating 
of the domain integral \eqref{eq:reduced_space_energy_ours}, because
the representation of the configuration, $\mathsf{q}$, is separated from the
representation of cubature. 

\section{Minimization via Cubature} 

\paragraph*{Minimization}
We minimize $E(\mathsf{q})$ using gradient descent~\cite{warp2022}. 
We initialize the increment at every cubature point with the explicit time stepping prediction 
$\Delta \bm{u}_i = h \bf{v}^j + h^2 M^{-1} f_{\textrm{ext}}$.
At every descent iteration, we compute the increments 
at all cubature points, and then find the best-fit increment to the latent configuration. The descent increment
at the $i$th cubature point is 
\begin{align} \label{eq:delta_x}
   \Delta \bm{u}_i = \alpha \left(\frac{M}{h^2}(\bm{\mathsf{W}}(\bm{X}_i)\mathsf{q}-\mathsf{q}_{\textrm{pred}}) 
   + \frac{\partial \Psi(\bm{X}_i+\bm{\mathsf{W}}(\bm{X}_i)\mathsf{q})}{\partial \bm{\mathsf{W}}(\bm{X}_i)\mathsf{q}} \right) \ .
\end{align}
After evaluating the full space increment at every cubature point, which we project to 
find the best fit subspace increment by minimizing the quadratic 
\begin{align}\label{eq:quadratic-projection}
\Delta \mathsf{q} &= \argmin_{\Delta \mathsf{q}}
\sum_i
w_i \big\|\bm{\mathsf{W}}(\bm{X}_i)\Delta \mathsf{q} - \Delta \bm{u}_i \big\|^2 \ , %
\end{align}
which amounts to solving a symmetric positive definite linear system.
The matrix depends only on the position and weight of the cubature points, 
and whilst these are invariant, 
a single Cholesky factorization allows for repeated
projections via backsubstitution.

When the cubature set changes, we reassemble the system matrix. 
Since $\bm{\mathsf{W}}(\bm{X}_i)$ is a function of $\bm{X}_i$, we
cache it at each cubature point, eliminating the network 
inference $\bm{\mathsf{W}}(\bm{X}_i)$ except
at newly introduced cubature points.

\paragraph*{Cubature Sampling}
Previous cubature sampling~\cite{An:Cubature:2008,Tycowicz:ECRDO:2013} provides promising results. One generates a set of training poses for the cubature optimization preprocess. This preprocess identifies desirable cubature points and associated nonnegative weights to achieve accurate energy approximation over the training poses.

But what about integrating subspace dynamics on novel meshes unseen during training? In this case,
the aforementioned approach is not directly applicable. 
We implemented a na{\"\i}ve cubature scheme, which we found satisfactory
for the examples that we tested. We
\begin{enumerate}
\item select $m$ vertices randomly from the tetrahedral mesh, and
\item additionally, select all the vertices incident to the $m$ vertices.
\end{enumerate}
These steps yield the equiweighted cubature points $\{\bm{X}_i\}$.

In all presented results that do not 
involve remeshing, we precompute the cubature scheme. 
For the remeshing examples, the cubature points in principle 
would change (locally) when the mesh is 
changed (locally). In our simplified demonstration of remeshing, 
where we know the sequence of meshes in advance, we precompute the cubature points 
and $\bm{\mathsf{W}}(\bf{X}_i)$ for all meshes.

\begin{table*}[t]
\caption{Performance statistics. We list the average simulation time cost (in seconds) for full-space simulations and reduced-space simulations. 
We also listed the number of sampled vertex and tetrahedrons.
A latent space dimension $r=20$ was used for all examples. The Young's modulus is $5\times 10^5$ for the dragon and bunny example, and is $2.5\times 10^6$ for other examples. We adopted a Poisson ratio of 0.25 for all examples.
Hardware: Intel Core i7-10750H.
}
\centering
\vspace*{-0.13in}
{\begin{tabular}{l |c c c c c c c c c c c}
\toprule
Example & vertex & tetrahedron & sampled vertex  & sampled tet  & full space & reduced space & speedup \\
& count & count & count & count ($\tilde{n}$) & step cost (ms) & step cost (ms) & \\
\midrule

Training with different shapes & 20K & 103K & 1.3K & 3K & 142 & 8   & 17 \\
Kinematic tearing & 20K & 91k-94K & 3.7K & 5.6K & 323 & 11 & 29   \\
Hole punching & 20K & 95K-100K & 3.8K & 6.3K & 288 & 13  & 22   \\
Falling Animals & 40K & 200k-210K & 1.3K & 2.1K & 350 & 8 & 43    \\ 
Interactive application & 100K & 516K & 1.4K & 2.2K & 335 & 6 & 56  \\ 
Dragon & 80K & 429K & 3.3K & 5K & 307 & 9 & 34  \\
Bunny & 20K & 101K & 1.6K & 2.4K & 267 & 9  &  29   \\

\bottomrule
\end{tabular}}

\label{tab:measurements}
\end{table*}

\vspace{0.15pt}
\begin{table*}[t]
\caption{\todoCY{
Statistics on precomputation. We include the data volume and time required for data generation and training for each example. During the data collection phase, we capture snapshots from various loading conditions, recording vertex displacements at specific time step. We also listed the total cost of all sampling operations, including selecting cubature points and caching $\bm{\mathsf{W}}(\bm{X})$.
} 
}
\centering

\vspace*{-0.13in}
{\begin{tabular}{l |c c c c c c c c c c c}
\toprule

Example & vertex & training snapshots  & number of & data generation  & training & cubature & evaluating \\
& count & count & loadings & cost (min) & cost (h) & selection (ms) & $\bm{\mathsf{W}}(\bm{X})$ (ms)\\
\midrule

Training with different shapes & 20K & 1650 & 5 & 3.4 & 4.9 & 15.9 & 32.7\\
Kinematic tearing & 20K  & 1300 & 2 & 24.0 & 4.1 & 90.6 & 210.2\\
Hole punching & 20K  & 5600 & 8 & 3.7 & 16.0  & 70.0 & 12.3 \\
Falling Animals & 40K & 3600 & 3 & 32.0 & 10.6 & 23.1 & 6.4 \\ 
Interactive application & 100K  & 1200 & 20 & 96.1 & 7.7 & 188.0 & 11.1 \\ 
Dragon & 80K  & 1275 & 1 & 13.3 & 5.1 & 40.6 & 14.1\\
Bunny & 20K  & 4800 & 8 & 12.9 & 13.4 & 26.2 & 10.4 \\
\bottomrule
\end{tabular}}

\vspace{-15pt}
\label{tab:training}
\end{table*}

\section{Results}
We conduct experiments to evaluate the unique features of
LiCROM. We ask whether
one neural basis, $(\bm{\mathsf{W}},\mathsf{q})$, can
\begin{enumerate}
\item be trained over diverse inputs generated by different meshes?
\item reproduce deformations on geometries seen during training?
\item and on novel geometries unseen during training?
\item facilitate mesh connectivity and topology changes?
\end{enumerate}

\subsection{Unique capabilities of a continuous ROM}

\paragraph*{Training with different shapes} 

\begin{figure}
\centering
\begin{tikzpicture}[x=0.5\textwidth, y=0.5\textwidth]

\node[anchor=south] (image) at (0,0) {
\includegraphics[width=8cm]{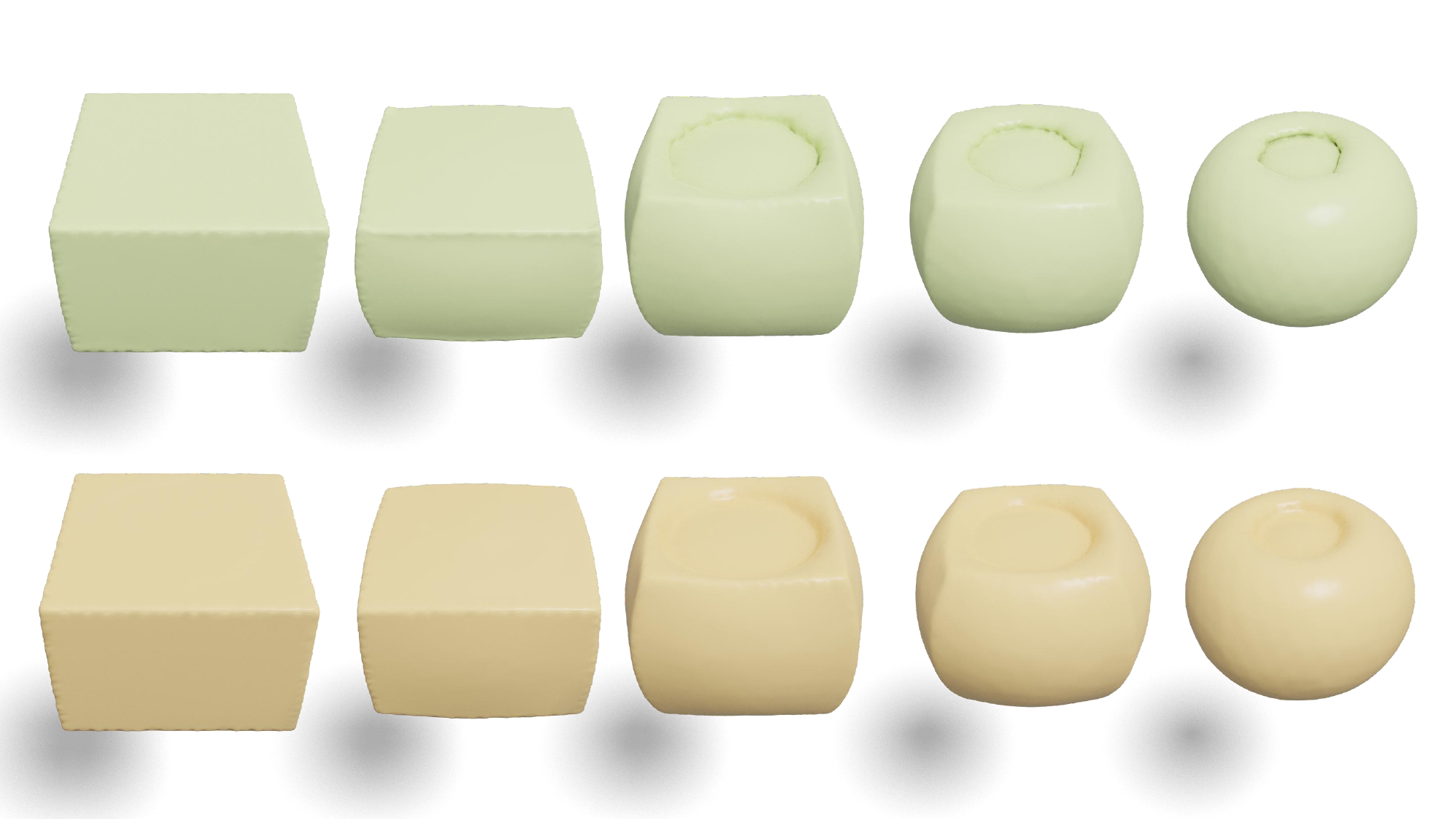}
};
\node at ([shift={(0, 0.51)}]image.south)[below] 
  {\footnotesize Training Data};
\node at ([shift={(0, 0.28)}]image.south)[below] 
  {\footnotesize Reduced Space Dynamics};  
\end{tikzpicture}

\vspace{-10pt}
\caption{
\emph{One neural basis spans the deformations of multiple shapes.}
During precomputation, we train one neural basis, $(\bm{\mathsf{W}},\mathsf{q})$,
with a single training set encompassing the full-space simulated
deformations of five shapes spanning cube to sphere (blue). Using
a continuous displacement field basis makes training on multiple shapes straightforward. During the online subspace dynamics, we simulate the same shapes with the same loading conditions (yellow),
observing good agreement, including for the top surface details. }\label{fig:cube2sphere}
\centering
\vspace{-15pt}
\end{figure}

We train one neural subspace $(\bm{\mathsf{W}},\mathsf{q})$
using a training set comprised of different shapes deformed under
similar load, and ask whether the subspace dynamics reconstruct the different behaviors of the shapes included in the training set.
We generate five shapes spanning cube to sphere, with equal bounding cubes, $[\pm0.5, \pm0.5, \pm0.5]$. We prescribe equal compressive displacement: for every vertex with underformed position near the top ($y<0.45$) or bottom ($y>-0.45$) we prescribe an equal downward ($-2m/s$) or stationary ($0m/s$)
velocity. A single training set includes the full-space dynamic deformations for these five meshes. We simulate the 
same five shapes in the 
reduced model (see Fig.~\ref{fig:cube2sphere}), noting 
agreement with the training data.

We repeat this experiment, this time with subspace dynamics
on novel shapes unseen during training (see Fig.~\ref{fig:TestInterpolation}). We observe good agreement
for the overall deformation, albeit sometimes with missing 
surface details, when these were not seen during training.

\begin{figure}
\centering
\begin{tikzpicture}[x=0.5\textwidth, y=0.5\textwidth]

\node[anchor=south] (image) at (0,0) {
\includegraphics[width=8cm]{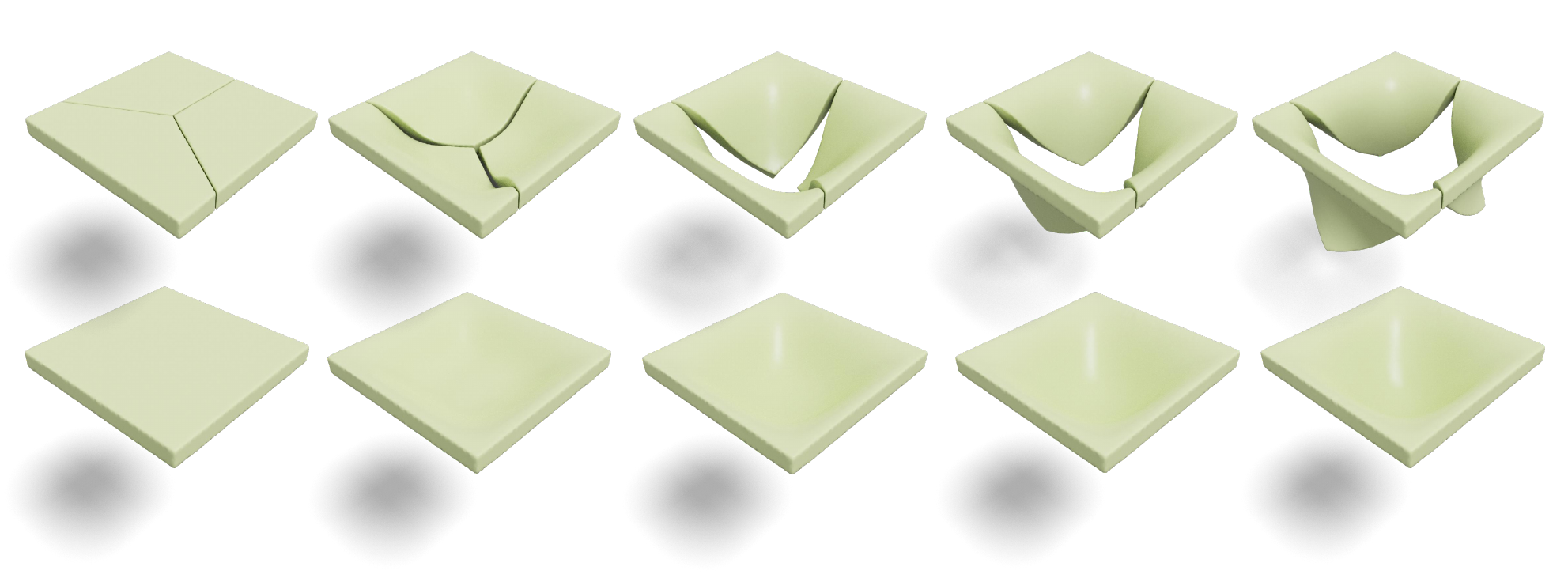}
};
\node at ([shift={(0, 0.07)}]image.south)[below] 
  {\footnotesize Training Data};
  
\end{tikzpicture}
\begin{tikzpicture}[x=0.5\textwidth, y=0.5\textwidth]

\node[anchor=south] (image) at (0,0) {

  \includegraphics[width=8cm]{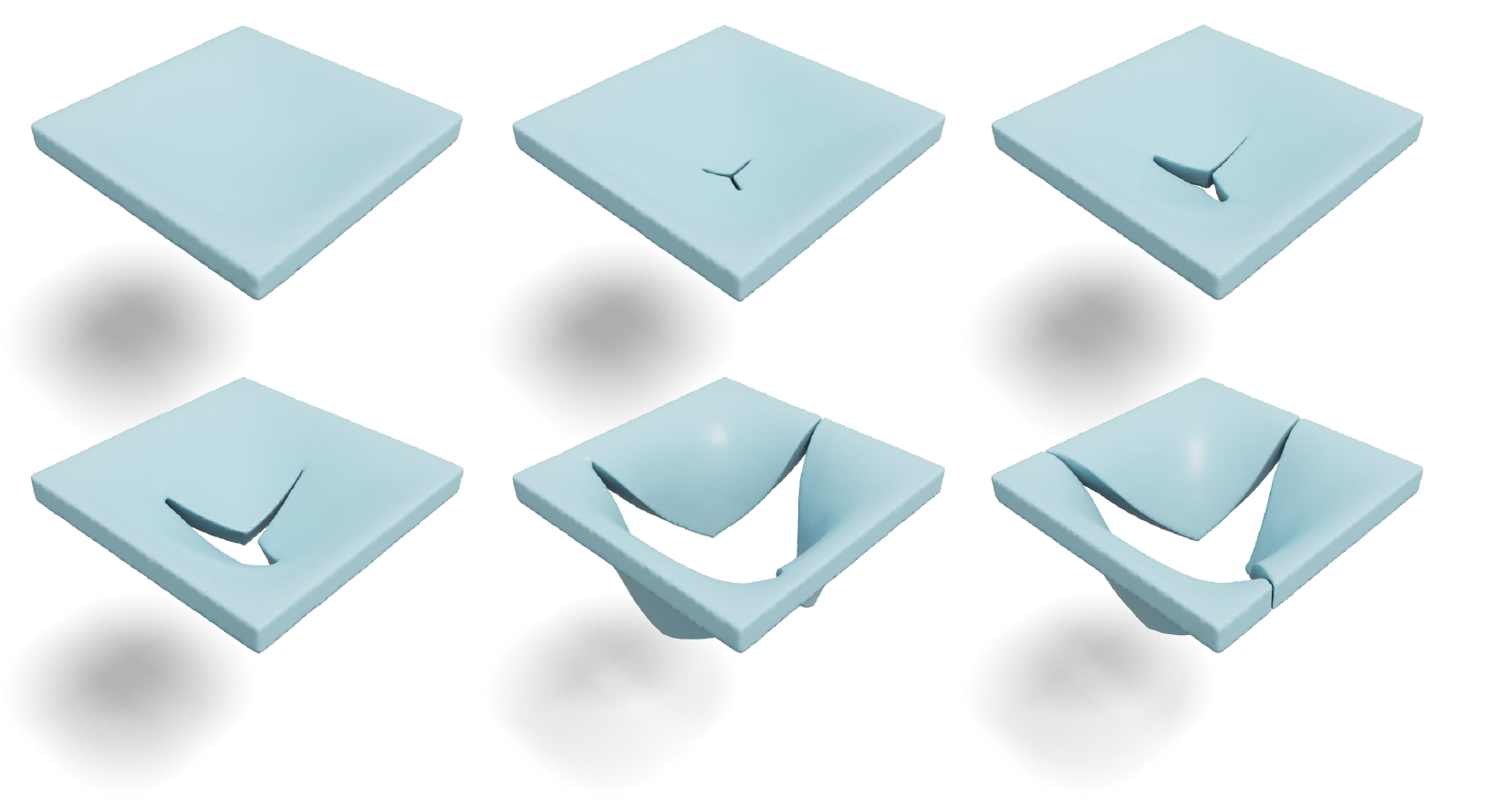}
};
\node at ([shift={(-0.18, 0.39)}]image.south)[below] 
  {$\rightarrow$};
  \node at ([shift={(-0.18, 0.37)}]image.south)[below] 
  {\footnotesize Cut};
  
\node at ([shift={(-0.18, 0.18)}]image.south)[below] 
  {$\rightarrow$};
\node at ([shift={(-0.18, 0.16)}]image.south)[below] 
  {\footnotesize Cut};
  
  \node at ([shift={(0.1, 0.39)}]image.south)[below] 
  {$\rightarrow$};
\node at ([shift={(0.1, 0.37)}]image.south)[below] 
  {\footnotesize Cut};

\node at ([shift={(0.1, 0.18)}]image.south)[below] 
  {$\rightarrow$};
  \node at ([shift={(0.1, 0.16)}]image.south)[below] 
  {\footnotesize Cut};

\node at ([shift={(0.39, 0.39)}]image.south)[below] 
  {$\rightarrow$};
  \node at ([shift={(0.39, 0.37)}]image.south)[below] 
  {\footnotesize Cut};
\node at ([shift={(0, 0.1)}]image.south)[below] 
  {\footnotesize Reduced Space Dynamics};
  
\end{tikzpicture}
\vspace{-10pt}
\caption{\emph{Kinematically-prescribed $\mathsf{Y}$-shaped tear.}
During precomputation, we train one neural basis on the 
full-space simulations of both an intact plate and a $\mathsf{Y}$- cut plate sagging under gravity. During the online subspace
dynamics, the plate is cut progressively (on a 
prescribed schedule) by redefining the connectivity of the cubature mesh. This novel partial-cut connectivity of the cubature mesh is unseen during training.
The deformations for the partial cut arise naturally from the available neural basis displacement
fields.
}\label{fig:fracture}
\centering
\vspace{-15pt}
\end{figure}
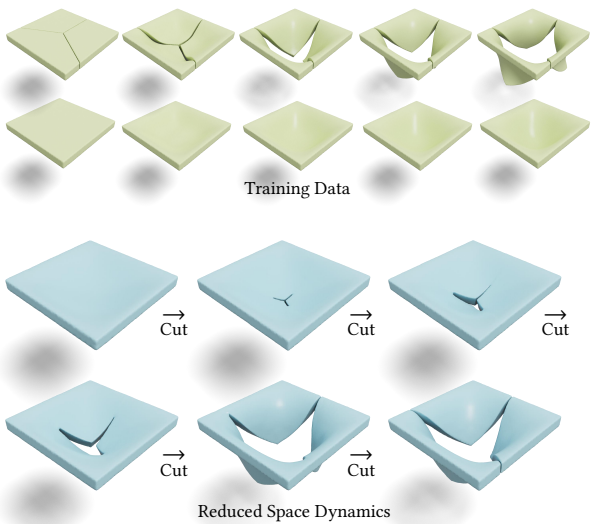

\paragraph*{Kinematic tearing}
Because the cubature mesh does not carry state,
it need not be tied to the meshing used 
in previous time steps nor the training phase. 
Combined with the ability to train on meshes
with different connectivity, these traits
make subspace modeling of tearing and fracture
easier (see Fig.~\ref{fig:fracture}).
During precomputation, we train one neural basis using a 
single training set comprised of two full-space simulations:
(1) a clamped plate sagging under gravity;
(2) the same plate, with a $\mathsf{Y}$-shaped cut, sagging under gravity.

In the online phase, we model the tearing of the plate
using subspace dynamics. Over time, we progressively redefine
the cubature mesh to grow a $\mathsf{Y}$-shaped cut (see Fig.~\ref{fig:fracture}). The cuts 
introduced in the cubature mesh have the desired effect
on the force computation, but they do not require a transfer
of state variables from the previous mesh.
Recall that the training set includes only the intact
and fully cut geometry; the deformations for the
partial cut arise (as in all linear subspace approaches)
from a weighted sum of the precomputed displacement fields.

A natural question then is ``how well does the continuous
neural displacement field capture a discontinuous deformation?'' 
This is particularly poignant as our implementation employs smooth ELU activation functions.
We visualize the basis displacement field $\bm{\mathsf{W}}(\bm{X})$
(see Fig.~\ref{fig:colorcode}), observing the discontinuity. 

Since the neural basis has no ``knowledge'' of the geometry, the cubature bears full responsibility for providing geometric
knowledge, and therefore producing distinct dynamics for distinct geometries. Undersampling produces artifacts (see Fig. \ref{fig:cubature}). Using $3713$ random samples (compared to
$20k$ vertices in the original data) is sufficient to 
obtain a $29\times$ speedup over the full space simulation.

\paragraph*{Hole punching}

In addition to simulating fractures, our method is capable of simulating the process of punching the cube and generating voids in real-time. In the example shown in Figure \ref{fig:PunchingCube}, we run simulation on 5 meshes with a fixed bottom under gravity. After training, we can simulate the process the cube being ``damaged'' (i.e. holes being cut out) by runtime remeshing. Note that after each remesh, the deformed position of the rest of the cube is consistent with the frame before the remeshing, except for the newly generated empty part. 

\paragraph*{Rolling animals}

Our method is able to simulate the collision and friction between the animals and the static inclined plane. For the example shown in Figure \ref{fig:FallingBunny}, when generating training data, we simulated an elastic animal under static gravity $g=-9.8m/s$. In each frame, we check if any vertex intersects with an infinite plane with normal $[0,\sqrt{2}/2, \sqrt{2}/2]$. If an intersection happens, we apply a penalty force along the normal of the plane to handle collision and set the velocity orthogonal to the plane normal to zero (infinite friction force). Results show that our latent space dynamics can reconstruct the colliding and rolling interaction between different animals and the plane.

\paragraph*{Animal interpolation}
After training on the three animals in Fig.~\ref{fig:animation-interpolation}, we interpolate among these three meshes via Wasserstein distances \citep{solomon2015convolutional}. Thanks to the discretization-agnostic nature of our method, we can readily deploy the previously trained model with all these meshes. Fig.~\ref{fig:animation-interpolation} demonstrates the corresponding latent space dynamics for each mesh.

\subsection{Interactive application}
We trained a neural basis on deformations induced by tugging at the armadillo
(see Fig.~\ref{fig:interactive}).
The full-space and reduced simulations require $335$ms and $6$ms per time step, 
respectively, on an Intel Core i7-10750. The $56\times$ speedup enables
interactive manipulation at $30$ frames per second.

The user can also load in previously unseen geometric models 
that can be swapped for the armadillo, mid-simulation,
without resetting the kinematic configuration or momentum. 
Note that the physical response is evaluated on the current geometry.
Although the kinematic training was conducted solely on the armadillo,
the physical response reflects the geometry, as evident, e.g.,
in the higher frequency oscillations of the thinner arms.
This demonstrates the one-shot generalization potential of LiCROM.
To the best of our knowledge, this is the first interactive-rate 
demonstration of model reduction that includes online 
substitutions of geometric model, including
previously unseen geometric models. Indeed, by training on a single geometry, our approach generalizes to other geometries, effectively achieving \emph{one-shot generalization}.

\subsection{Comparison with nonlinear CROM}

Our method shares a similar motivation with Continuous Reduced-Order Modeling (CROM) \citep{CHEN:CROM-MPM:2023,CHEN:CROM:2023}. Both seek discretization independence. In CROM, a nonlinear decoder $(\mathsf{q},\bm{X}) \mapsto \bm{u}$ maps
the reduced configuration and reference position to the corresponding deformed position.
Compared to a linear basis, a nonlinear approach may be more complex to implement, analyze and 
compute, or more carefully chosen training data  to avoid overfitting.
We observed artifacts when applying CROM to deformable simulation, which spurred
our investigation into a linear subspace (see Fig.~\ref{fig:compare_CROM}).  

LiCROM offers an important
advantage over CROM in the projection \eqref{eq:quadratic-projection},
which, due to the linearity of the basis, becomes a simple
minimization of a quadratic, i.e., the solution of a linear system
which, by prefactorization, can be reused along with cubature points.
By contrast, the nonlinearity of the CROM basis~\cite{CHEN:CROM:2023} 
does not allow for such a trivial projection. 

We leverage this
fast projection (which amounts to just backsubstitution on the prefactored
matrix) to implement implicit time stepping, which requires
\emph{repeated} projections each time step. In the nonlinear CROM,
each such projection would require multiple
expensive network Jacobian evaluations.

\section{Discussion}

We have presented the first discretization-independent linear model reduction method, in the sense that the subspace basis does not 
explicitly store, refer, or rely on particulars of the 
discretizations employed to generate the training set,
integrate the forces, or output the resulting animation.

This discretization-independence is achieved by
defining the subspace basis vectors as \emph{continuous}
displacement fields over the reference domain, which
we implement using neural implicit fields.

Consequently, we are able to demonstrate that a single 
subspace model can be trained from differing discretizations
or even differing geometries. The learned basis can accelerate simulation 
by about 20--50$\times$ whilst supporting phenomena not typically seen
in subspace methods, such as phenomena that typically require
remeshing (e.g., cutting), changes to topology (e.g., hole punching), or novel geometry unseen during training.

\paragraph*{Limitations}
These novel features are accompanied by novel limitations.
First, the trained subspace is of course limited by the observed
data. For a neural implicit field, this usual limitation is accompanied by a novel aspect: the field will not hesitate to
``hallucinate'' an extrapolated result in portions of the reference domain 
$\Omega$ that had few or no data observations. 
As a corollary, if we train a displacement basis on a thin geometry, this basis
may not be suitable for a thick geometry, where some cubature points will 
sample a potentially unsuitable extrapolation of the displacement field. It
would be interesting to incorporate regularizers for such extrapolation~\cite{Liu:Lipschitz:2022}. \todoCY{Fig.\ref{fig:Failure} (a) and (b) demonstrate two modes of generalization failures of our approach: vastly different loading conditions and geometric sizes.}

\begin{figure}
\centering
\subfigure[Illustration of test data]{
\begin{tikzpicture}[x=0.5\textwidth, y=0.5\textwidth]
\node[anchor=south] (image) at (0,0) {
  \includegraphics[width=8cm]{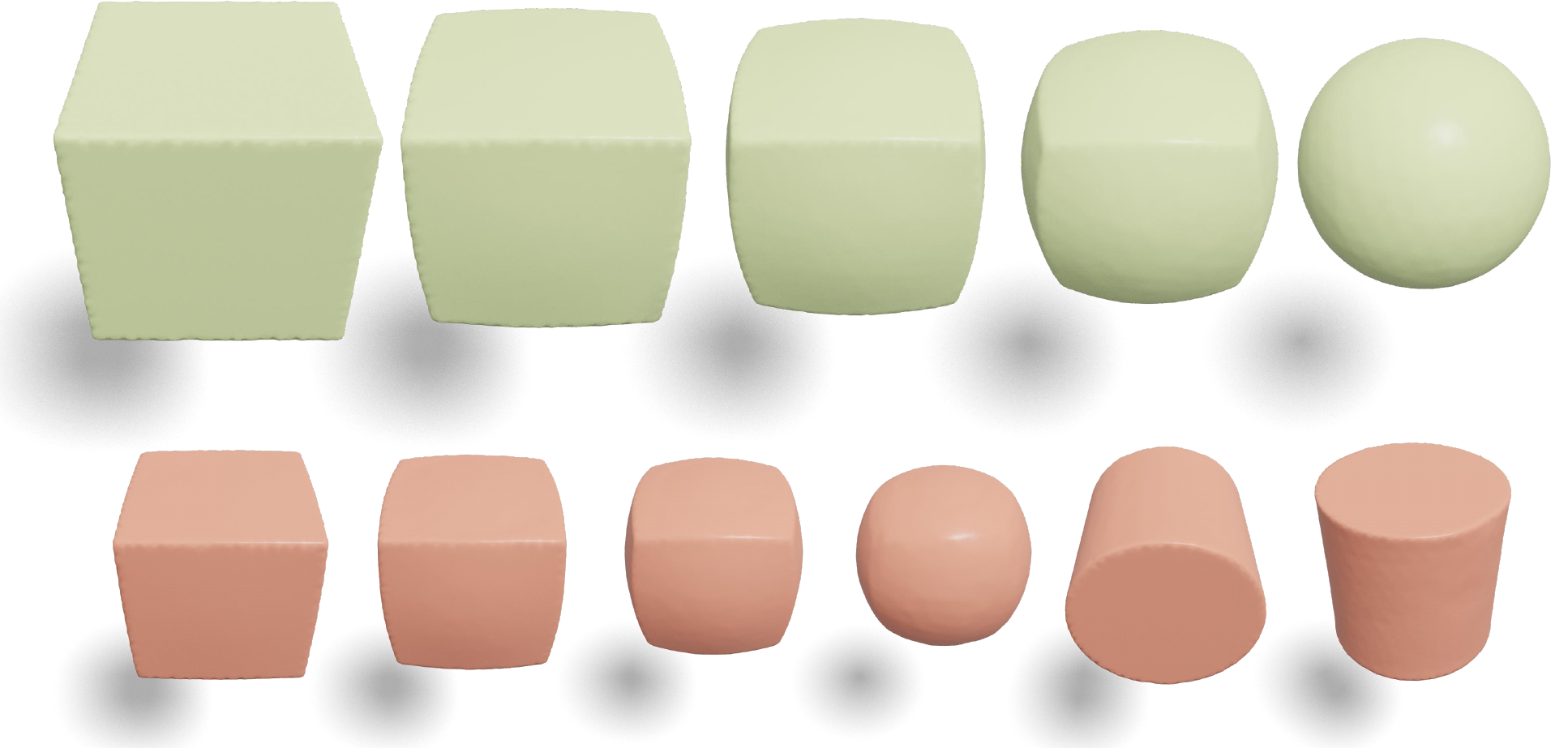}
};
\node at ([shift={(0.0, 0.5)}]image.south)[below] 
  {\footnotesize Shapes in Training Data};
\node at ([shift={(0.0,0.23)}]image.south)[below] 
  {\footnotesize Test Shapes, not in Training Data};
\end{tikzpicture}
}
\subfigure[Latent space dynamics]{
\begin{tikzpicture}[x=0.5\textwidth, y=0.5\textwidth]
\node[anchor=south] (image) at (0,0) {
  \includegraphics[width=8cm]{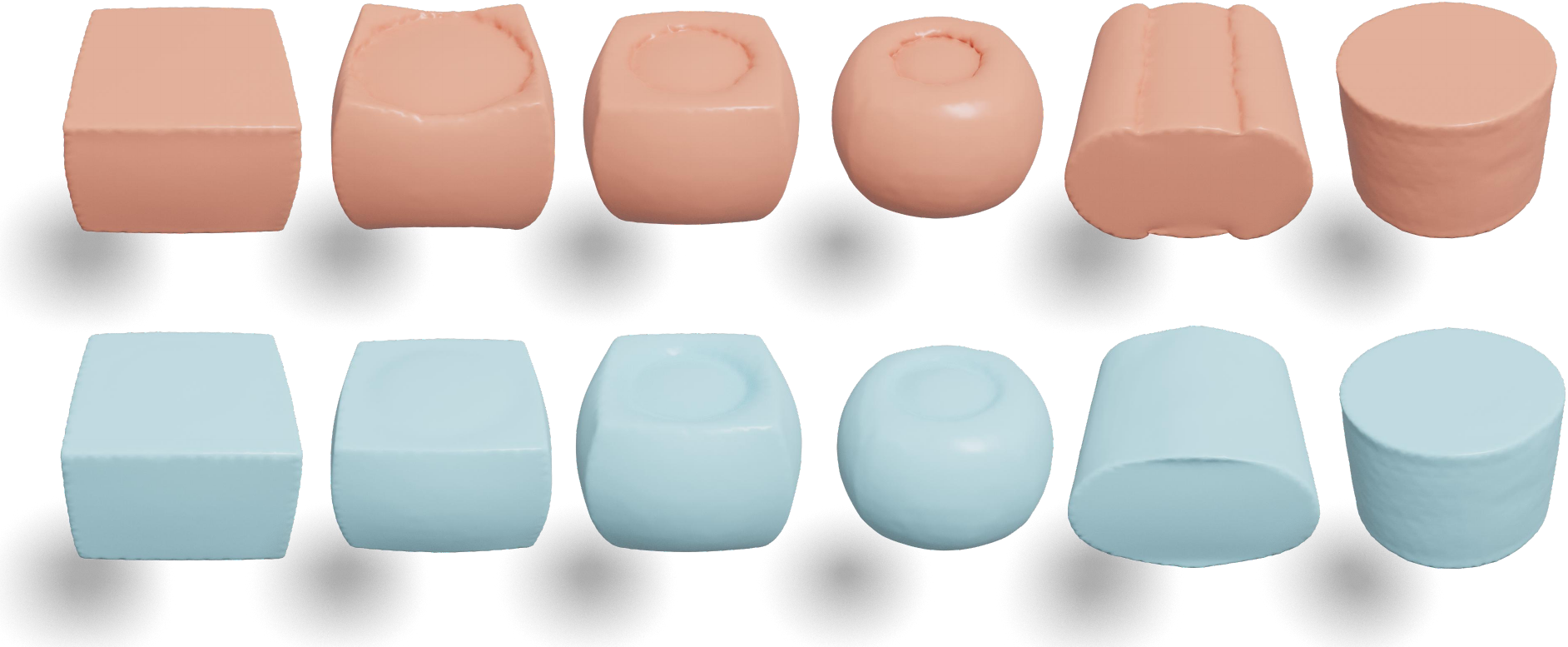}
};
\node at ([shift={(0.0, 0.43)}]image.south)[below] 
  {\footnotesize Full Space Dynamics, not in Training Data};
\node at ([shift={(0.0,0.24)}]image.south)[below] 
  {\footnotesize Reduced Space Dynamics};
\end{tikzpicture}
}
\caption{
\emph{One neural basis can span plausible deformations for new shapes.}
During precomputation, we train one neural basis, $(\bm{\mathsf{W}},\mathsf{q})$,
with a single training set encompassing the full-space simulated
deformations of five shapes spanning cube to sphere (blue).
The training omits the test shapes (\todoCY{coral}). 
During the online subspace dynamics, we simulate the test shapes with the same loading conditions,
observing good general agreement, albeit with 
some missing surface details unseen during training. 
}
\label{fig:TestInterpolation}
\end{figure}

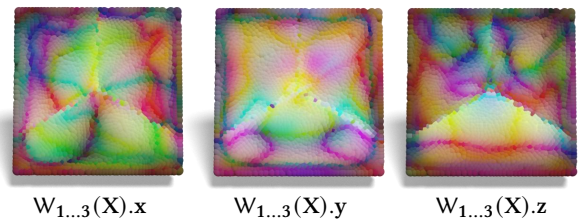
\begin{figure}
\centering
    
\begin{tikzpicture}[x=0.5\textwidth, y=0.5\textwidth]

\node[anchor=south] (image) at (0,0) {
\includegraphics[width=8cm]{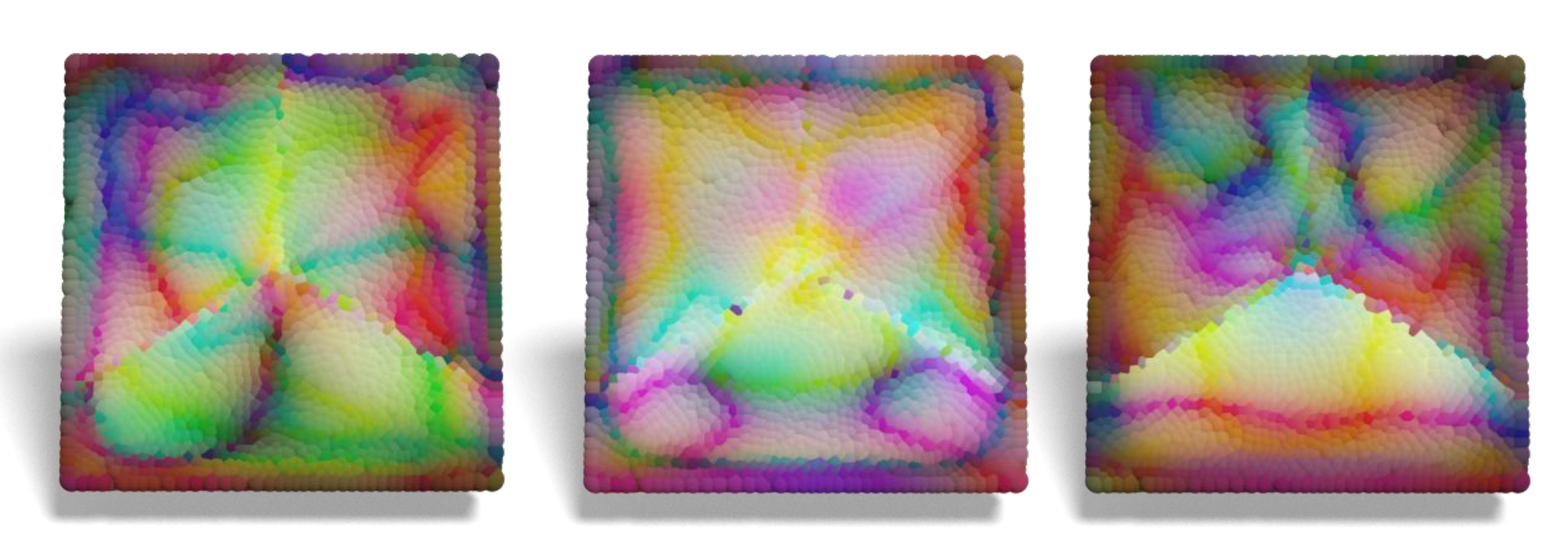}
};

\node at ([shift={(-0.3,0.03)}]image.south)[below] {$\bf{\mathsf{W}}_{1\ldots 3}(\bm{X}).x$};

\node at ([shift={(0,0.03)}]image.south)[below] {$\bf{\mathsf{W}}_{1\ldots 3}(\bm{X}).y$};

\node at ([shift={(+0.3,0.03)}]image.south)[below] {$\bf{\mathsf{W}}_{1\ldots 3}(\bm{X}).z$};

\end{tikzpicture}

\caption{\emph{Displacement basis visualization}. 
We visualize the first three dimensions of the continuous reduced basis $\bm{\mathsf{W}}(\bm{X})$ 
over the reference domain $\Omega$ for the tearing scenario (see Fig.~\ref{fig:fracture}). 
Red, green, and blue correspond to the displacements of $\bm{\mathsf{W}}_1(\bm{X})$, $\bm{\mathsf{W}}_2(\bm{X})$ and $\bm{\mathsf{W}}_3(\bm{X})$, respectively.
Each $\bm{\mathsf{W}}_i(\bm{X})$ is vector-valued: we visualize
the $x$, $y$, and $z$ components of displacement in the left, middle, and right, respectively.
It is evident that the basis includes displacement discontinuities, particularly 
along the $x z$ plane of the plate.
}\label{fig:colorcode}
\centering
\end{figure}

\begin{figure}
\centering
\begin{tikzpicture}[x=0.5\textwidth, y=0.5\textwidth]
\node[anchor=south] (image) at (0,0) {
  \includegraphics[width=6cm]{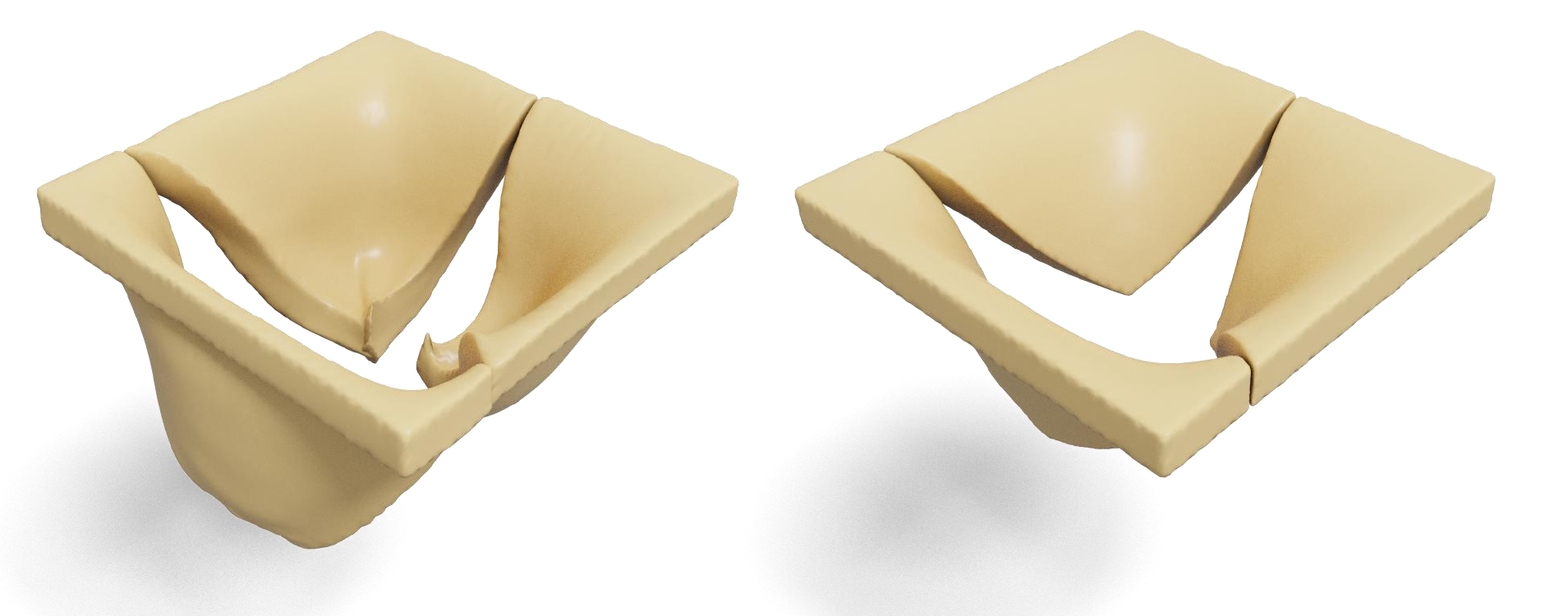}
};
\node at ([shift={(-0.15,0.03)}]image.south)[below] 
  {\footnotesize 376 sampled vertices};
\node at ([shift={(+0.15,0.03)}]image.south)[below] 
  {\footnotesize 3713 sampled vertices};
\end{tikzpicture}
\caption{\emph{Comparison of cubature point density} 
for the tearing scenario (see Fig.~\ref{fig:fracture}), comparing 376 (left) versus 3713 (right) sampled vertices.\label{fig:cubature}}
\centering
\end{figure}

\begin{figure}
\centering
\begin{tikzpicture}[x=0.5\textwidth, y=0.5\textwidth]

\node[anchor=south] (image) at (0,0) {
 \includegraphics[width=8cm]{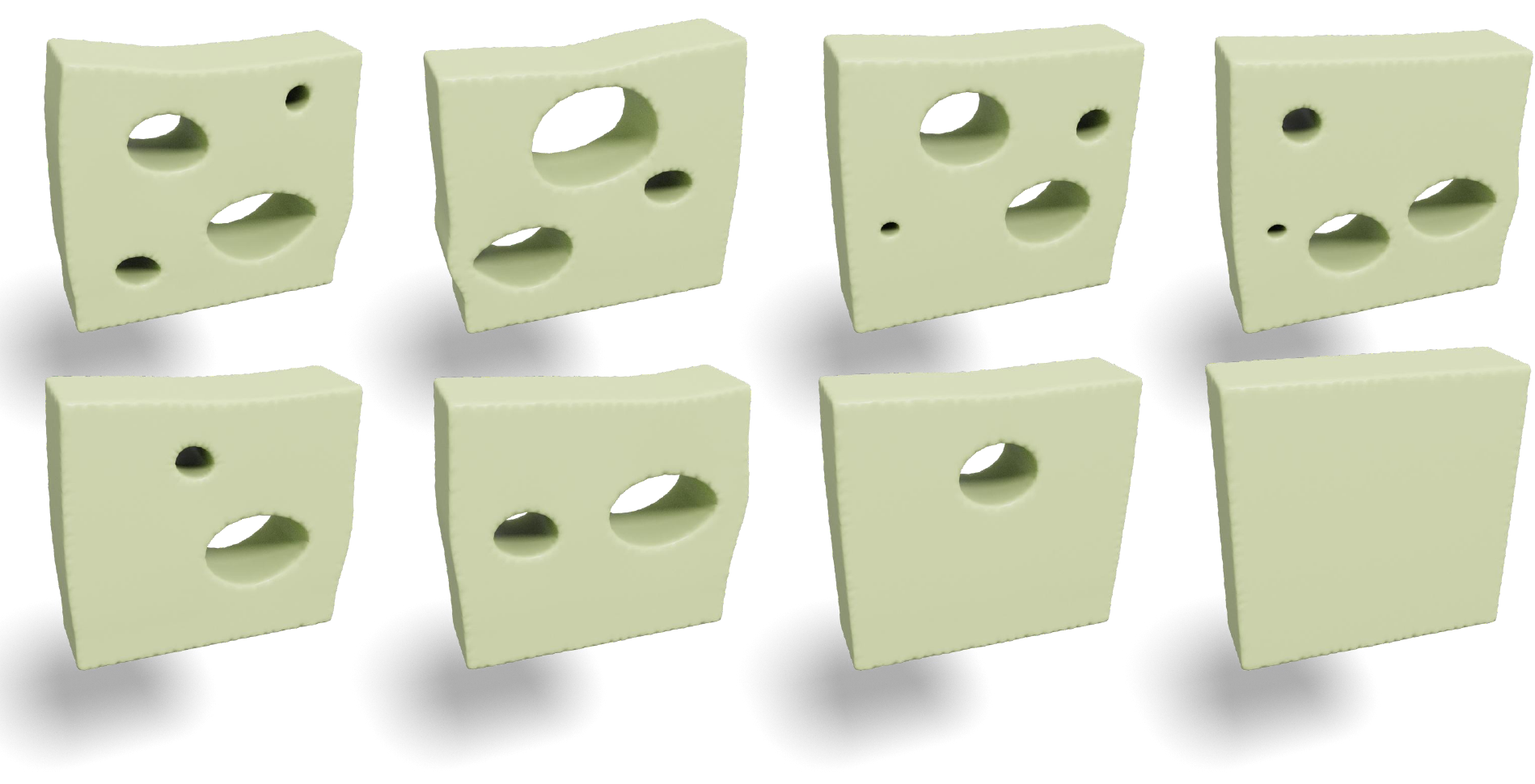}
};

  \node at ([shift={(0, 0.06)}]image.south)[below] 
  {\footnotesize Training Data};  
\end{tikzpicture}

\begin{tikzpicture}[x=0.5\textwidth, y=0.5\textwidth]

\node[anchor=south] (image) at (0,0) {
\includegraphics[width=8cm]{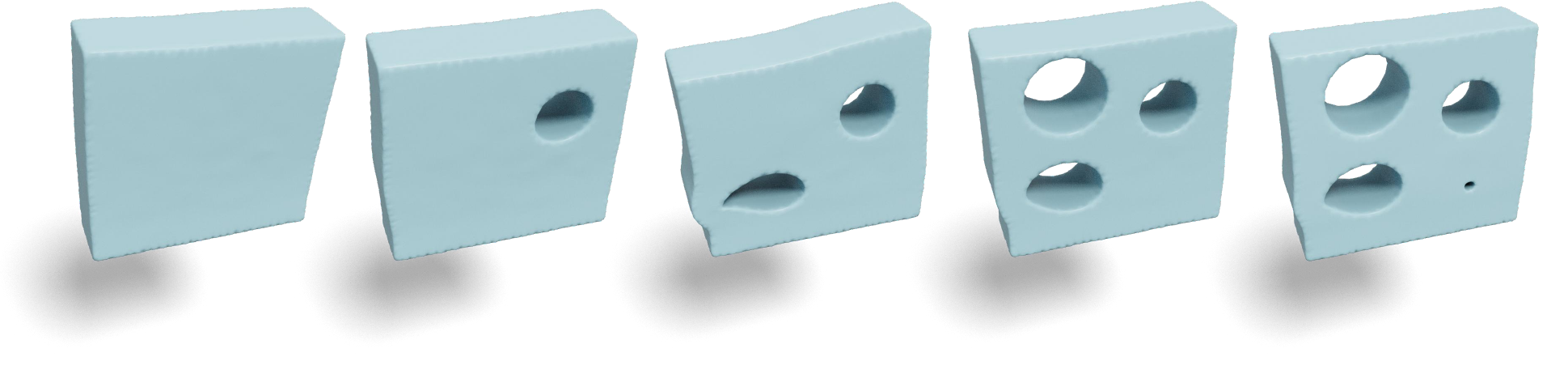}
};
\node at ([shift={(-0.26, 0.095)}]image.south)[below] 
  {$\rightarrow$};
\node at ([shift={(-0.08, 0.095)}]image.south)[below] 
  {$\rightarrow$};  
\node at ([shift={(0.1, 0.095)}]image.south)[below] 
  {$\rightarrow$};
\node at ([shift={(0.28, 0.095)}]image.south)[below] 
  {$\rightarrow$};  

\node at ([shift={(-0.26, 0.065)}]image.south)[below] 
  {\footnotesize Punch!};
\node at ([shift={(-0.08, 0.065)}]image.south)[below] 
  {\footnotesize Punch!};  
\node at ([shift={(0.1, 0.065)}]image.south)[below] 
  {\footnotesize Punch!};
\node at ([shift={(0.28, 0.065)}]image.south)[below] 
  {\footnotesize Punch!};

\node at ([shift={(0, 0.01)}]image.south)[below] 
  {\footnotesize Reduced Space Dynamics};  
\end{tikzpicture}

\caption{Punching Cube. We train this example using 5 meshes with different void. After we tain a network with the 5 sequence, we are able to simulate the process of punching the cube and generate voids at runtime. In latent space dynamics, we apply remeshing whenever we "punch" a new void.}\label{fig:PunchingCube}
\centering
\end{figure}
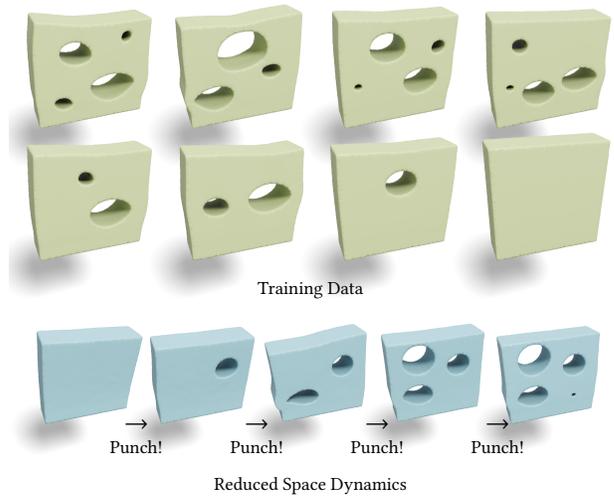

\begin{figure}
\centering
\begin{tikzpicture}[x=0.5\textwidth, y=0.5\textwidth]
\node[anchor=south] (image) at (0,0) {
 \includegraphics[width=6cm]{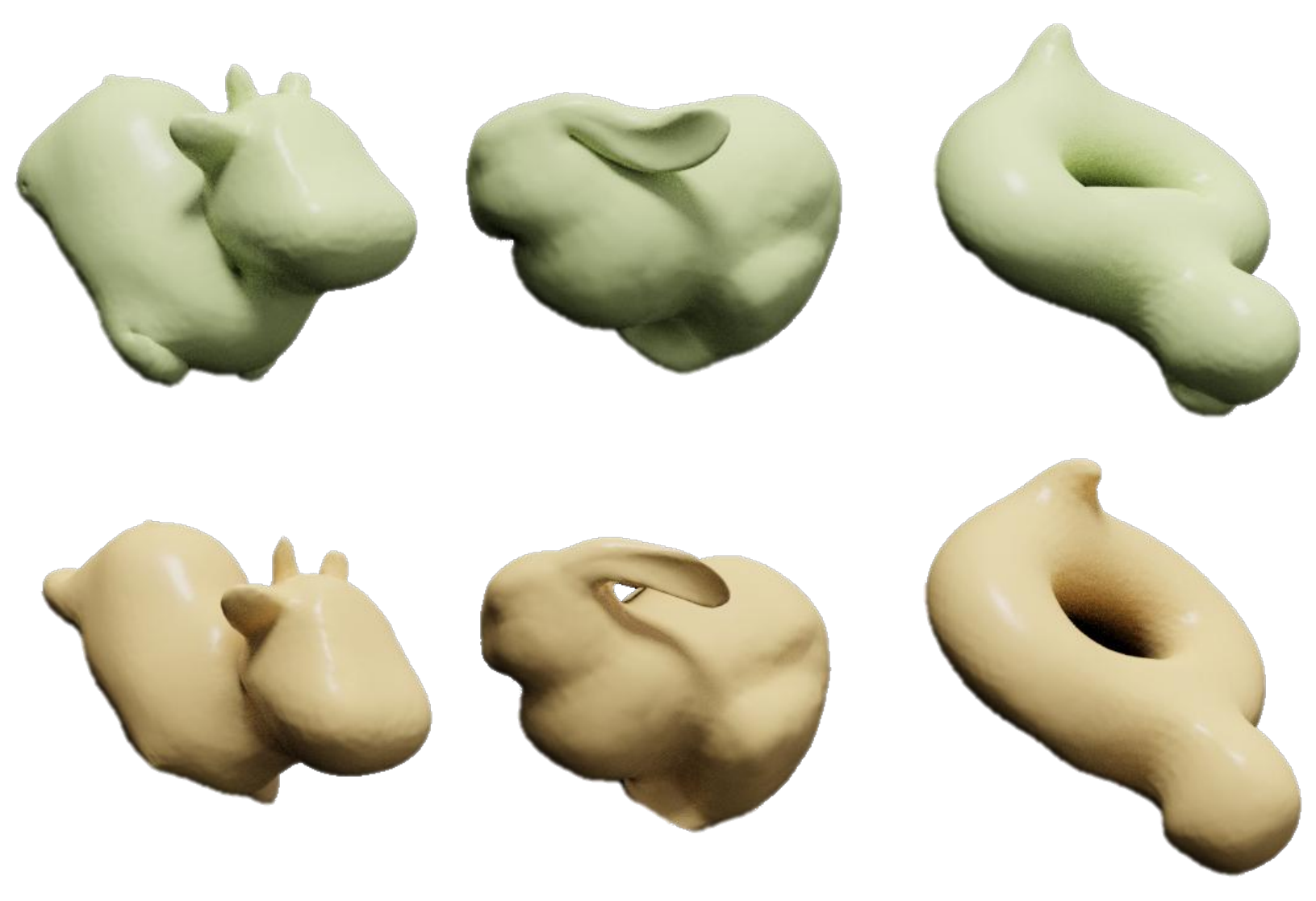}
};
\node at ([shift={(0.0, 0.28)}]image.south)[below] 
  {\footnotesize Training Data};
\node at ([shift={(0.0,0.04)}]image.south)[below] 
  {\footnotesize Reduced Space Dynamics};
\end{tikzpicture}
\caption{\emph{Rollin' along}. We learn a subspace for the deformation induced by
rolling of the bunny, the duck ``bob'', and the cow ``spot'' down an inclined plane. 
We then apply this subspace to simulate the deformations of myriad 
interpolated animal shapes.
}\label{fig:FallingBunny}
\centering
\end{figure}

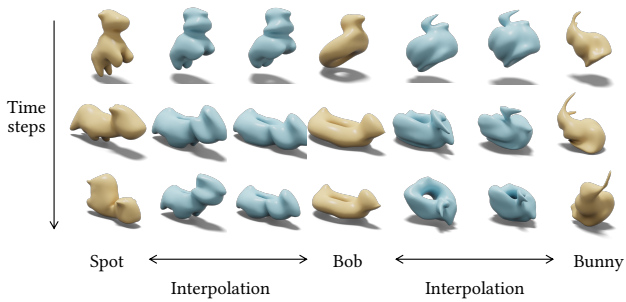
\begin{figure}
\centering
\begin{tikzpicture}[x=0.5\textwidth, y=0.5\textwidth]
\node[anchor=south] (image) at (0,0) {
   \includegraphics[width=\linewidth]{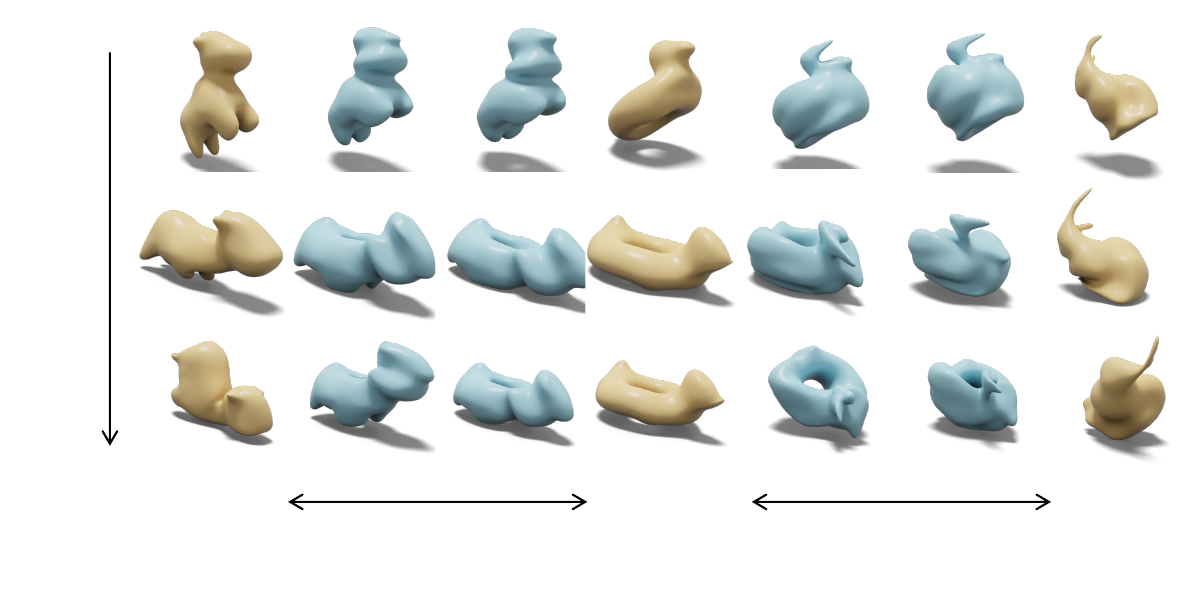}
    };
 \node at ([shift={(-0.43, 0.34)}]image.south)[below] 
  {\footnotesize Time}; 
\node at ([shift={(-0.43, 0.31)}]image.south)[below] 
  {\footnotesize steps}; 

\node at ([shift={(-0.31, 0.11)}]image.south)[below] 
  {\footnotesize Spot}; 
\node at ([shift={(0.05, 0.11)}]image.south)[below] 
  {\footnotesize Bob}; 
\node at ([shift={(0.425, 0.11)}]image.south)[below] 
  {\footnotesize Bunny}; 

  \node at ([shift={(-0.14, 0.07)}]image.south)[below] 
  {\footnotesize Interpolation}; 
\node at ([shift={(0.24, 0.07)}]image.south)[below] 
  {\footnotesize Interpolation}; 
  
\end{tikzpicture}
\vspace{-0.8cm}
\caption{\emph{Animation interpolation}. After training on the dynamics of a few meshes, our method can simulate the dynamics of the interpolated meshes.
}\label{fig:animation-interpolation}
\centering
\end{figure}

Indeed, since the training of the subspace has no explicit knowledge about the geometry, the trained subspace may fail to reconstruct certain surface details when tested on novel geometry that is not included in the training data, as shown in Fig.~\ref{fig:TestInterpolation}(b). It would be interesting to ameliorate this limitation by introducing an explicit ``geometry code'' when training and later using the network.

Second, the combined training of a neural implicit field and PointNet is expensive compared to POD, requiring several hours. This is the cost we trade for the benefit of PointNet's permutation
invariance. Interestingly, if this permutation invariance were discarded in lieu of a simpler,
permutation-dependent decoder, some aspects of discretization-independence would remain.
In particular, while the resulting embedding would no longer be independent of 
input discretization, the resulting displacement field basis would \emph{still} be continuous
and therefore \emph{not} impose any discretization on
the cubature scheme nor the subspace dynamics output.
Future work may involve accelerating training while
retaining permutation-invariance.

Our model also shares the shortcomings and benefits of linear-subspace model reduction methods: the dimension of the subspace typically exceeds that of nonlinear approaches, regardless of whether the displacement-field is encoded as a discrete ~\cite{Fulton:LSD:2018} or continuous~\cite{CHEN:CROM:2023,CHEN:CROM-MPM:2023} field. However---with the exception of methods developed in the computational math community \cite{Lee2018rom}---the state of the art in nonlinear approaches (especially in graphics) still
seems to rely on linear subspaces for regularization~\cite{Fulton:LSD:2018,Shen:HOD:2021}; 
perhaps these same kinds of regularizations can be applied in the continuous
domain, e.g., by regularizing CROM with LiCROM.

Unlike other linear-subspace ROMs, ours is not trained using POD, nor does
the training objective explicitly ask for orthogonality.
Orthogonality optimizes the conditioning of the basis, and is desirable
for reducing error during projection; we did not observe any challenges with projection.
We intend to evaluate the angle between basis vectors and report this in the near future.

\paragraph*{Future work} 
Our preliminary implementation leaves open many immediate steps.  
We employed a random cubature sampling approach with equal weights, 
solely for its simplicity and immediacy. Recall that the
$\mathsf{Y}$-shaped tear required $3.7k$ random samples. 
It seems reasonable to
expect that a data-aware sampling approach 
in the spirit of~\citeN{An:Cubature:2008} could
reduce the number of cubature points. Since our examples 
include geometries unseen during training, the sampling strategy
would have to be adapted to the data at runtime.

Following \emph{warp}, we used gradient descent to minimize the energy, however, alternatives abound.
For instance, our implementation is immediately
amenable to incorporating an (L-)BFGS solver, which approximates
Newton's method without using a Hessian. Indeed,
due to the linearity of the subspace, computing the reduced energy Hessian, as required for an exact
Newton's method, is straightforward via 
$\hess_{\mbox{$\mathsf{q}$}} 
\Psi(\bm{X}+\bm{u}(\mbox{$\mathsf{q}$})) = 
{\bm{\mathsf{W}}(\bm{X})^T}\,
{\hess_{\bm{u}}} \Psi(\bm{X}+\bm{\mathsf{W}}(\bm{X})
\mbox{$\mathsf{q}$}) \, 
{\bm{\mathsf{W}}(\bm{X})},$
which can be assembled via cubature at $\{\bm{X}_i\}$.
Note that the exact Hessian evaluation does not require 
differentiating through the neural network, 
which would be the case for a 
nonlinear subspace~\cite{Fulton:LSD:2018,CHEN:CROM:2023}.

Although we began with \emph{warp} on the GPU, 
we ultimately implemented our online subspace dynamics
solely on the CPU with \emph{taichi}.
In our intended application domains (virtual reality, games) 
there is significant contention over GPU acceleration, which
is primarily reserved for rendering. Achieving interactive
rates on a CPU, albeit more limiting, was an important criterion.
However, for other use cases, a fast GPU implementation 
remains desirable, and we intend to re-implement this method
on the GPU.

\paragraph{Open source}
Our implementation of LiCROM will be released.

\begin{acks}
We would like to thank Otman Benchekroun, Jonathan Panuelos, Kateryna Starovoit, and Mengfei Liu for their feedback on Fig 1. We would also like to thank our lab system administrator, John Hancock, and our financial officer, Xuan Dam, for their invaluable administrative support in making this research possible. This project is funded in part by Meta and the Natural Sciences and Engineering Research Council of Canada (Discovery RGPIN-2021-03733). We thank the developers and community behind PyTorch, the Taichi programming language, and NVIDIA Warp for empowering this research. The meshes in Fig \ref{fig:interactive} are derived from entries 133568, 133078 and 170179 of the Thingi10k dataset ~\cite{Thingi10K}.
\end{acks}

\bibliographystyle{ACM-Reference-Format}
\bibliography{sample-bibliography}

\clearpage

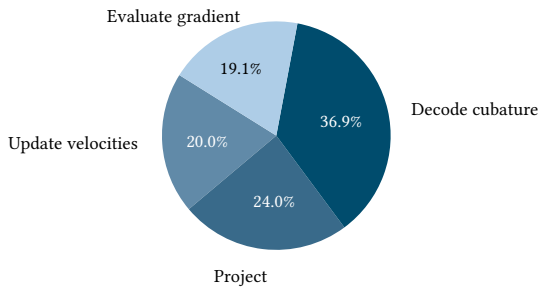
\begin{figure}
    \centering

    \begin{tikzpicture}[x=0.5\textwidth, y=0.5\textwidth]

    \node[anchor=south] (image) at (0,0) {
    \includegraphics[width=0.7\linewidth]{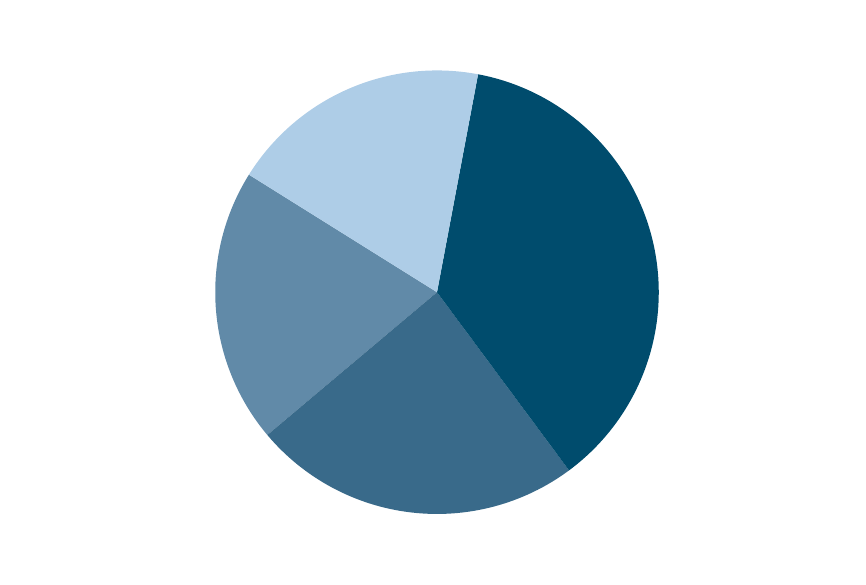}
    };

  \node at ([shift={(-0.3, 0.25)}]image.south)[below] 
  {\footnotesize Update velocities}; 
  \node at ([shift={(-0.1, 0.25)}]image.south)[below] 
  {\footnotesize \textcolor{white} {20.0\%}}; 

\node at ([shift={(0.3, 0.3)}]image.south)[below] 
  {\footnotesize Decode cubature}; 
\node at ([shift={(0.1, 0.28)}]image.south)[below] 
  {\footnotesize \textcolor{white} {36.9\%}}; 

\node at ([shift={(-0.05, 0.05)}]image.south)[below] 
  {\footnotesize Project}; 
\node at ([shift={(-0.0, 0.16)}]image.south)[below] 
  {\footnotesize \textcolor{white} {24.0\%}}; 

\node at ([shift={(-0.15, 0.44)}]image.south)[below] 
  {\footnotesize Evaluate gradient}; 
\node at ([shift={(-0.05, 0.36)}]image.south)[below] 
  {\footnotesize  {19.1\%}}; 

\end{tikzpicture}

    \vspace{-0.3cm}
    
    \caption{\emph{Breakdown of computational cost} for interactive manipulation (see Fig.~\ref{fig:interactive}). In operation order: 
    Over all cubature points, (a) \textbf{update velocities} by evaluating $\bm{\mathsf{W}}(\bm{X})\dot{\mathsf{q}}$; (b) \textbf{decode cubature} by evaluating
    $\bm{\mathsf{W}}(\bm{X})\mathsf{q}$; (c) \textbf{evaluate the gradient} 
    $\eqref{eq:delta_x}$. Finally, (d)
    \textbf{project} to the reduced space by least squares $\eqref{eq:quadratic-projection}$,
    solving the prefactored linear system using backsubstitution.
    }
    \label{fig:runtime_breakdown}
\end{figure}

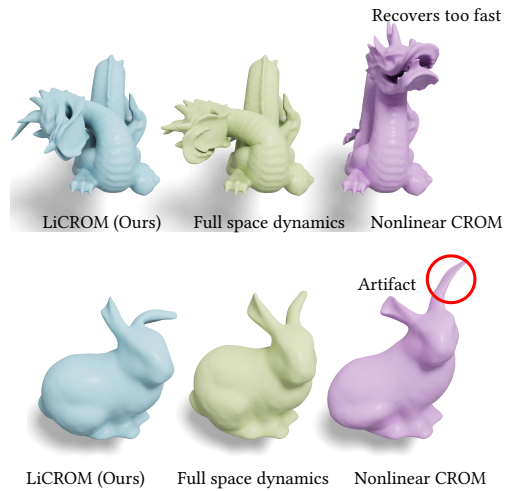
\begin{figure}
\centering

\begin{tikzpicture}[x=0.5\textwidth, y=0.5\textwidth]
\node[anchor=south] (image) at (0,0) {
  \includegraphics[width=6cm]{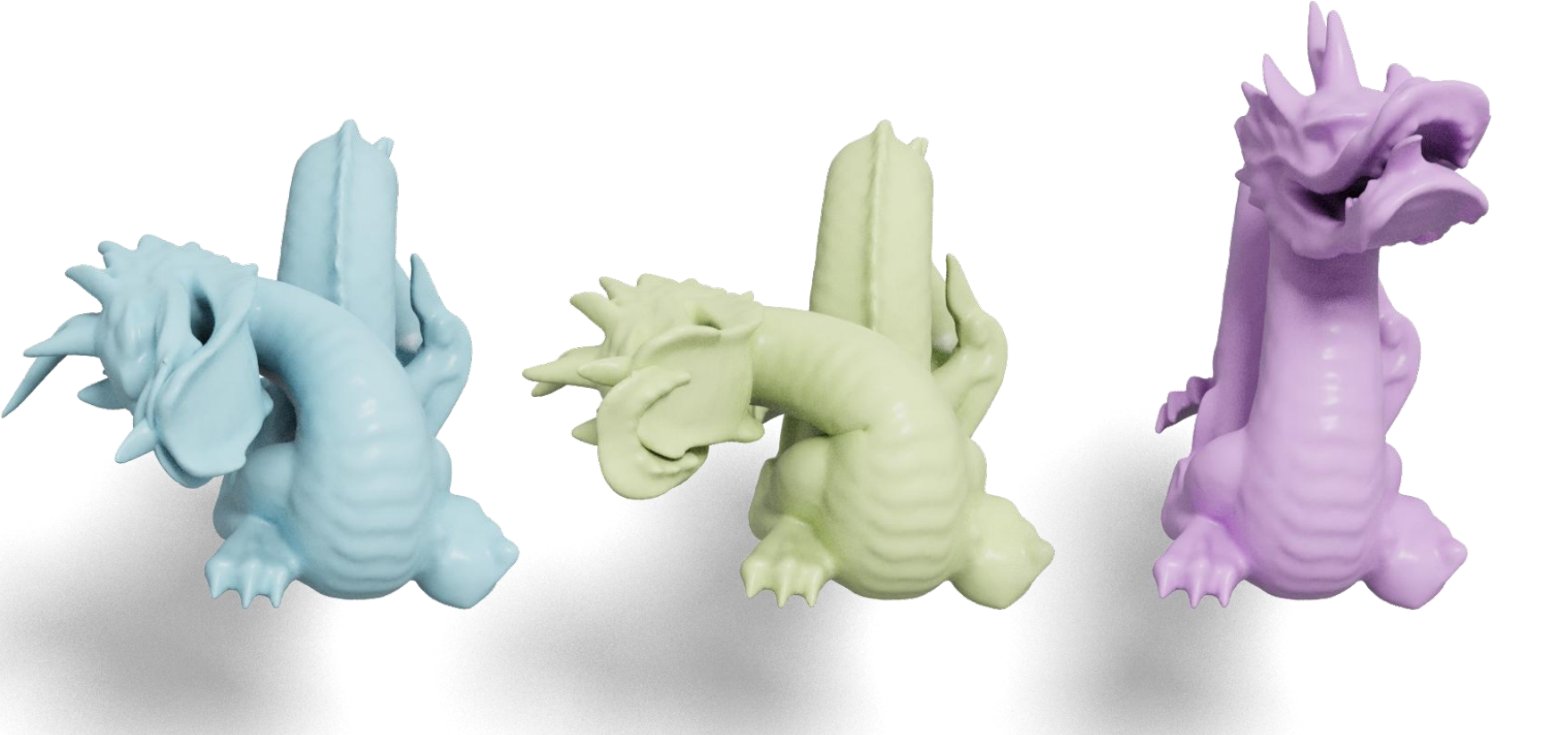}
};
\node at ([shift={(-0.2, 0.05)}]image.south)[below] 
  {\footnotesize LiCROM (Ours)};
\node at ([shift={(0.05,0.05)}]image.south)[below] 
  {\footnotesize Full space dynamics};
\node at ([shift={(0.3,0.05)}]image.south)[below] 
  {\footnotesize Nonlinear CROM};

\node at ([shift={(0.3,0.36)}]image.south)[below] 
  {\footnotesize Recovers too fast};

\end{tikzpicture}

\begin{tikzpicture}[x=0.5\textwidth, y=0.5\textwidth]
\node[anchor=south] (image) at (0,0) {
  \includegraphics[width=6cm]{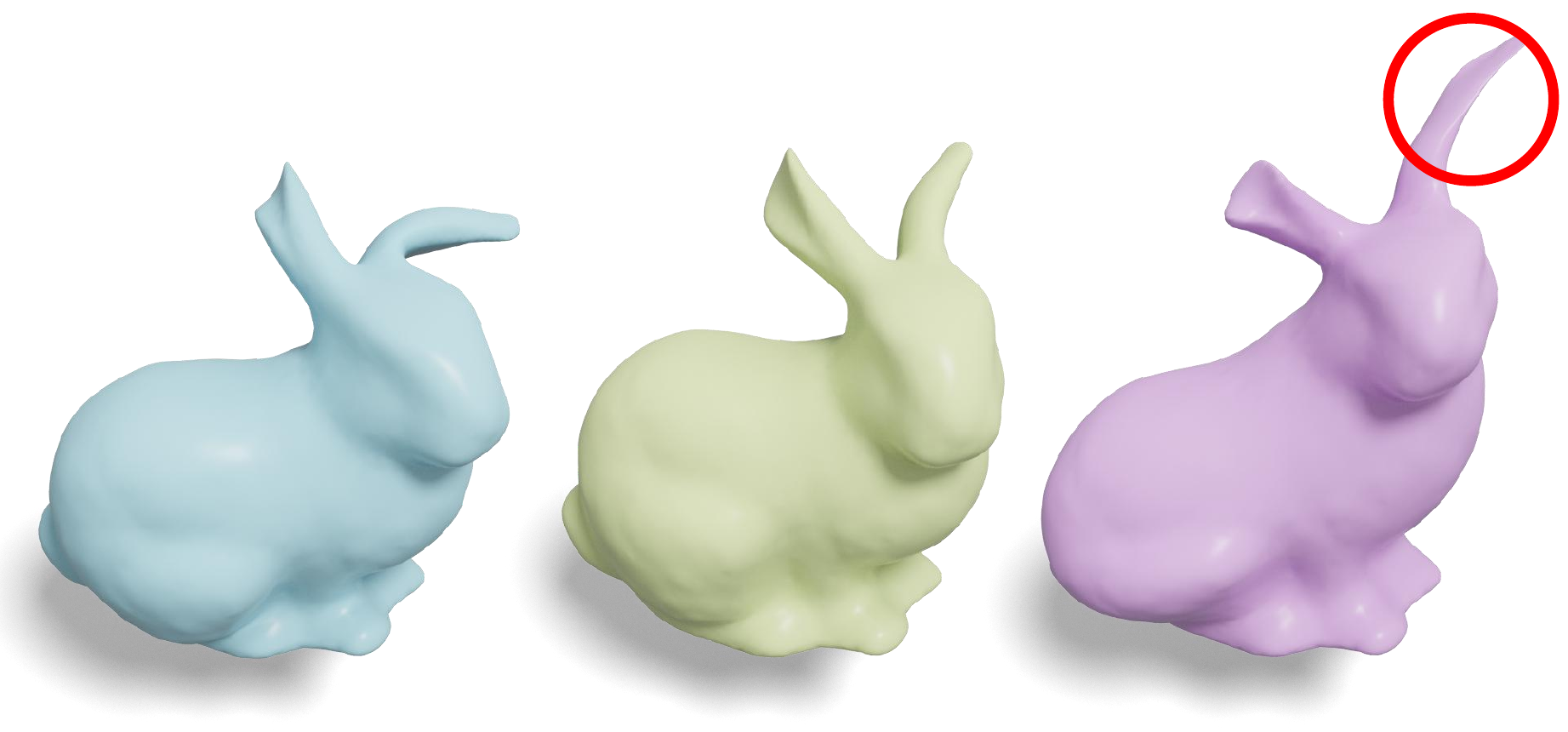}
};
\node at ([shift={(-0.25, 0.01)}]image.south)[below] 
  {\footnotesize LiCROM (Ours)};
\node at ([shift={(0.0,0.01)}]image.south)[below] 
  {\footnotesize Full space dynamics};
\node at ([shift={(0.25,0.01)}]image.south)[below] 
  {\footnotesize Nonlinear CROM};

\node at ([shift={(0.2,0.3)}]image.south)[below] 
  {\footnotesize Artifact};
\end{tikzpicture}

\vspace{-0.4cm}
\caption{Comparison with CROM. From left to right: our method, ground truth full-space simulation, CROM. In the dragon's example, we applied a force on the dragon's head that is not included in training data. We found that CROM suffers from overfitting and looks less similar to ground truth compared with our method. In the bunny's example, there are visible artifacts on the bunny's ear using CROM, while our method generates a reasonable result.}\label{fig:compare_CROM}
\centering
\end{figure}

\begin{figure}
\centering
\subfigure[Generalization over loading conditions (failure case).]{

\begin{tikzpicture}[x=0.5\textwidth, y=0.5\textwidth]

\node[anchor=south] (image) at (0,0) {
\includegraphics[width=8cm]{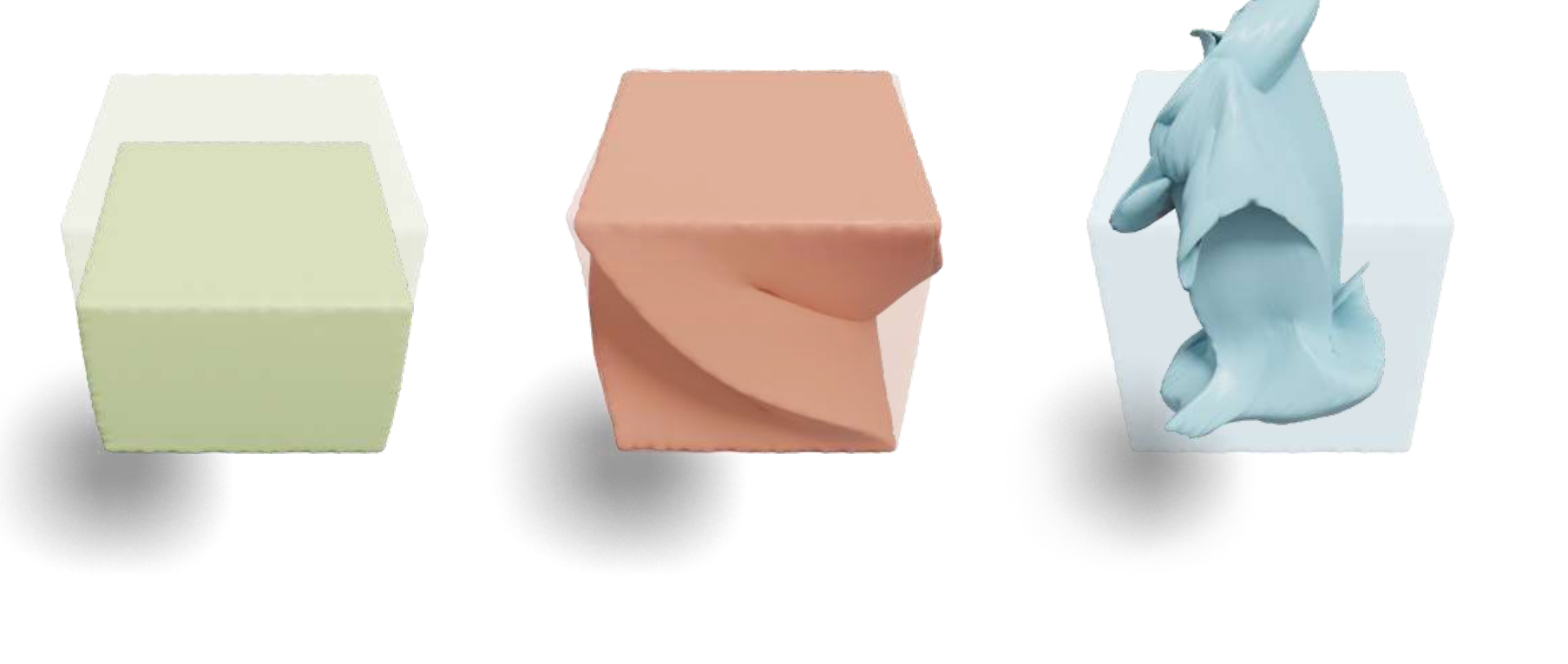}
};

  \node at ([shift={(-0.31, 0.1)}]image.south)[below] 
  {\footnotesize Training Data};

  \node at ([shift={(-0.01, 0.1)}]image.south)[below] 
  {\footnotesize Full Space Dynamics};
  \node at ([shift={(-0.01, 0.065)}]image.south)[below] 
  {\footnotesize (Not in Training Data)};

  \node at ([shift={(0.28, 0.1)}]image.south)[below] 
  {\footnotesize Reduced Space Dynamics};
  \node at ([shift={(0.28, 0.065)}]image.south)[below] 
  {\footnotesize (Not in Training Data)};
  
\end{tikzpicture}

}
\vspace{-0.3cm}

\subfigure[Generalization over rescaling of shape (failure case).]{

\begin{tikzpicture}[x=0.5\textwidth, y=0.5\textwidth]

\node[anchor=south] (image) at (0,0) {
\includegraphics[width=\linewidth]{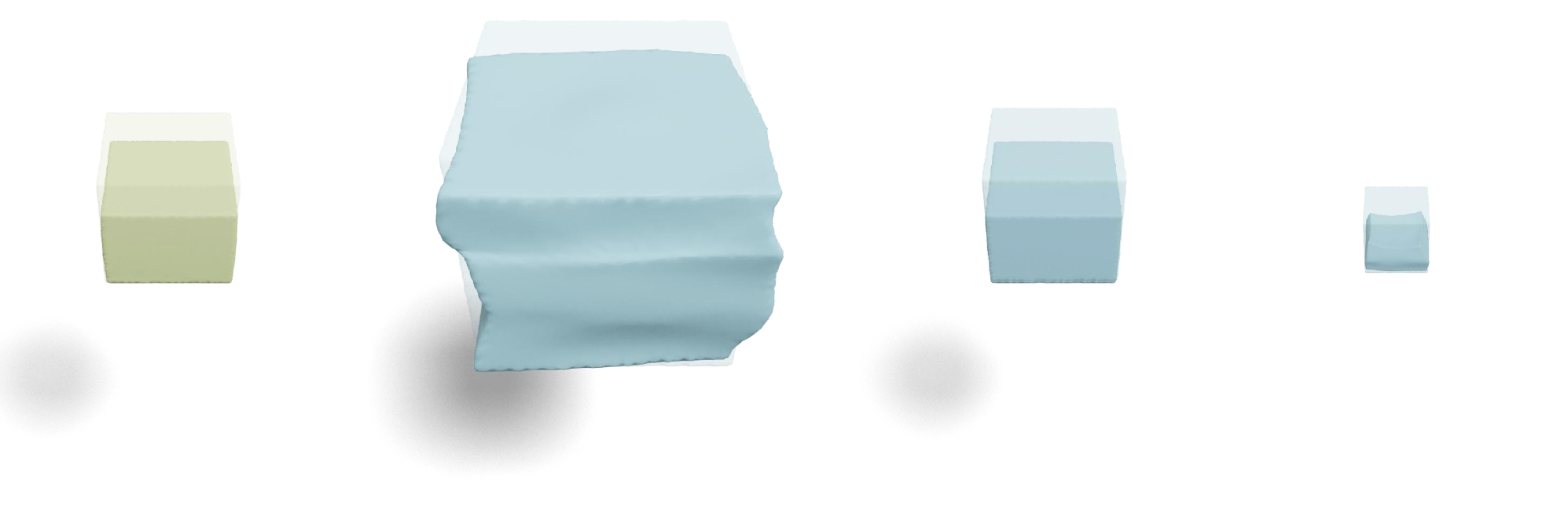}
};

\node at ([shift={(-0.38, 0.05)}]image.south)[below] 
  {\footnotesize Training data};

\node at ([shift={(-0.12, 0.05)}]image.south)[below] 
  {\footnotesize Scale 2x};

\node at ([shift={(0.17, 0.05)}]image.south)[below] 
  {\footnotesize Scale 1x};

 \node at ([shift={(0.37, 0.05)}]image.south)[below] 
  {\footnotesize Scale 0.5x};
  
\end{tikzpicture}

}
\vspace{-0.2cm}
\caption{
\todoCY{
Failures of generalization: We tested the displacement field from Fig. \ref{fig:cube2sphere} (training with different shapes). (a) We consider generalization over loading conditions. 
Training on a cube compressed by a vertical load and testing on a cube 
subject to rotational loads reveals poor agreement between full and reduced space dynamics.
(b) We also consider generalization over spatial scale. Training on a unit cube with vertical load,  
and testing on cubes where both spatial extent and loading are scaled accordingly, 
we observe unexpected surface features, particularly when scaling up.}
}
\vspace*{-0.12in}
\label{fig:Failure}
\end{figure}
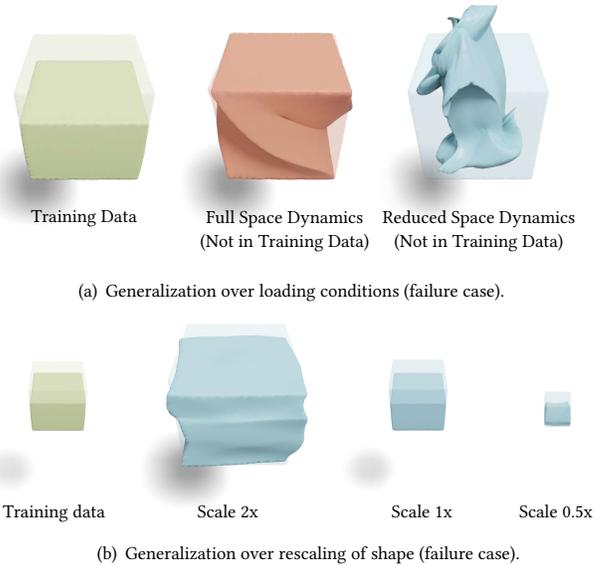

\cleardoublepage

\end{document}